\documentclass[%
 reprint,
 amsmath,amssymb,
 aps,
prc,
]{revtex4-1}

\usepackage{graphicx}
\usepackage{dcolumn}
\usepackage{bm}
\usepackage{amsmath}
\usepackage{slashed}
\usepackage{rotating}
\usepackage{float}
\usepackage{color}
\usepackage[normalem]{ulem}

\newcommand{\Lagr}{\mathcal{L}}
\newcommand{\no}{\nonumber}
\newcommand{\ba}{\begin{eqnarray}}
\newcommand{\ea}{\end{eqnarray}}
\def\bra#1{\left\langle #1\right|}
\def\ket#1{\left| #1\right\rangle}

\begin{document}

\preprint{APS/123-QED}

\title{$S=-1$ meson-baryon interaction and the role of isospin filtering processes}

\author{A. Feijoo}
 \affiliation{Nuclear Physics Institute, 25068 \v Re\v z, Czech Republic}

\author{V. Magas}
\author{A. Ramos}%
\affiliation{
 Departament de F\' isica Qu\`antica i Astrof\' isica and Institut de Ci\`encies del Cosmos, Universitat de Barcelona, Mart\' i Franqu\`es 1, E08028 Barcelona, Spain.}

\date{\today}

\begin{abstract}
A study of the meson-baryon interaction in the $S=-1$ sector is performed, employing a chiral SU(3) Lagrangian up to next-to-leading order (NLO) and implementing unitarization in coupled channels. The model is constrained by a large set of experimental data, paying special attention to processes that are sensitive to the NLO contributions, such as  the $K^- p\to K^+\Xi^-, K^0\Xi^0$ reactions. The consideration of additional cross sections in single isospin channels, $K^-p\to \eta\Lambda, \eta\Sigma$, has been found to provide NLO low-energy constants of rather similar size. The stability of these constants has also been tested by the inclusion of 
explicit resonant terms. 
Predictions for new isospin filtering processes, like the $I=1$ $K^0_L p \to K^+ \Xi^0$ reaction that could be measured at the proposed secondary $K^0_L$ beam at Jlab, or the weak decay of the $\Lambda_b$  into a $J/\Psi$ and different meson-baryon pairs in $I=0$, available at LHCb, are presented. The measurement of such reactions would put valuable constraints on the chiral models describing the $S=-1$ meson-baryon interaction.
\end{abstract}

\maketitle

\section{Introduction}
\label{Intro}

Unitaritzed chiral perturbation theory (UChPT) has shown as an appropriate framework to treat the low-energy meson-baryon interaction in the $S=-1$ sector. This non-perturbative scheme is based on the inherited symmetries from quantum chromodynamics (QCD), particularly the (spontaneously broken) chiral symmetry, and the unitarity and analyticity of the scattering amplitude. The combination of these two guiding principles permits not only a suitable reproduction of the experimental data but also the dynamic generation of bound-states and resonances (see \cite{ollerreport} and references therein).

The description of the $\Lambda(1405)$ resonance in terms of a molecular state arising from coupled-channel meson-baryon re-scattering is a paradigmatic case of success in this sector. Its dynamical origin was predicted in the late 1950s \cite{L1405} and, almost forty years later, it was reformulated in terms of the modern chiral effective theory approach \cite{KSW}, after which several studies were devoted to the $\bar{K} N$ interaction with different degrees of sophistication \cite{KWW,OR,OM,LK,BMW,GarciaRecio:2002td,BFMS,BNW}. The developed models reproduced the $\bar{K} N$ scattering data satisfactorily and pinned the  $\Lambda(1405)$ down as a superposition of two poles of the scattering amplitude, generated dynamically from the unitarization in coupled channels \cite{OM,2pole,PRL}. Despite these theoretical breakthroughs, and as it was immediately pointed out by the authors of \cite {BMN}, the precise location of these poles required additional subthreshold information on the antikaon-nucleon dynamics.

In the last years, several experimental groups have carried out measurements that shed some new light on this topic. The COSY Collaboration at J\"ulich \cite{Zychor:2007gf} and the HADES Collaboration at GSI \cite{Agakishiev:2012xk} provided invariant $\pi\Sigma$ mass distributions from $pp$ scattering experiments; while mass distributions of $\Sigma^+ \pi^-$, $\Sigma^- \pi^+$, and $\Sigma^0 \pi^0$ states in the region of the $\Lambda(1405)$ \cite{Moriya:2013eb}, as well as differential cross sections \cite{Moriya:2013hwg} and a direct determination of the expected spin-parity $J^\pi=1/2^-$ of the $\Lambda(1405)$ \cite{Moriya:2014kpv} have been measured by the CLAS Collaboration at JLAB. However, the most relevant experimental data that should be considered are the precise energy shift and width of the $1s$ state in kaonic hydrogen by the SIDDHARTA Collaboration \cite{SIDD} at DA$\Phi$NE. This experimental achievement establishes the strong $K^-p$  scattering length up to a precision of around $20$\% and, therefore, settles the dispute between the DEAR \cite{DEAR} and KEK \cite{Iwasaki:1997wf} measurements with almost a factor $2$ as relative uncertainty. After these new experimental data became available, this topic has experienced a renewed interest. The theoretical models have been revisited \cite{IHW,HJ_rev,GO,Mizutani:2012gy,Roca:2013av,Roca:2013cca,Mai:2014xna} as a response to the need to extend the approach to higher orders and energies aiming at greater accuracy in data description and to determine better the properties of the $\Lambda(1405)$. 

 It should be noted that the fits carried out to develop these models were accommodated to the two-body cross sections of $K^- p$ scattering into $\pi\Sigma, \bar{K}N,\pi\Lambda$ states; besides, the authors of \cite{Roca:2013av,Roca:2013cca,Mai:2014xna} incorporated experimental photoproduction data on $\gamma p \to K^+ \pi \Sigma$ reactions to extract information about the two-pole structure of the $\Lambda(1405)$ resonance. From \cite{OR,OM,LK,BMW,GarciaRecio:2002td,2pole,BFMS,BNW,BMN,IHW,Mizutani:2012gy,Mai:2014xna}, it can be concluded that the significant term that allows one to get a good agreement with the experimental data is the Weinberg-Tomozawa (WT) one, which is of order $O(p)$, and the addition of other terms such as the direct and crossed Born terms ($O(p)$) as well as the next-to-leading order (NLO) terms ($O(p^2)$) play merely a fine-tuning role. The only exception is the case of \cite{GO}, where authors performed a fit which included, apart from the classical channels, scattering data from $K^- p \to \eta \Lambda , \pi^0 \pi^0 \Sigma^0 $ and data from two event distribution ($K^- p \to \Sigma^+(1660) \pi^- \to \Sigma^+ \pi^- \pi^+ \pi^-$ and $ K^- p \to \pi^0 \pi^0 \Sigma^0$). The difference lies on the fact that they obtained a notable improvement in reproducing the $K^- p \to \eta \Lambda$ reaction once the NLO contribution was taken into account. 
 
Another challenging aspect derived from the above theoretical framework is the determination of the non constrained low energy constants involved in the NLO terms of the chiral Lagrangian. The dissimilarity in the values of the NLO coefficients found by the former studies made us turn our attention to processes in which the higher order terms of the Lagrangian could play a significant role beyond that of fine tuning.  A clear example of such processes are the $K^- p \to K \Xi$ reactions since they receive no direct contribution from the Weinberg-Tomozawa (WT) term and the rescattering terms due to the coupled channels are not sufficient to reproduce the experimental scattering data. In \cite{Feijoo:2015yja}, we included the $ K\Xi $ scattering data into the fitting procedure to finally demonstrate the sensitivity of the $K \Xi$ channels to the NLO term without deteriorating the quality of the data description. Despite the evidence of a non-negligible s-wave contribution of the Born terms in \cite{OM}, which reaches $\sim 20$\% of the dominant WT contribution at $1.5$~GeV, the Born terms were assumed to be very moderate based on the results of  \cite{BNW,Mizutani:2012gy,IHW} which show therein that the inclusion of them led to tiny changes in the fitting parameters and the quality of the fits was barely affected. 

As a matter of fact, the relevance of the Born diagrams of the chiral model in the $\bar{K}N \to K^+\Xi^-, K^0 \Xi^0$ cross sections would not have come as a surprise if one would have considered the work \cite{Jackson:2015dva}, which was published almost at the same time as \cite{Feijoo:2015yja}. The authors of the former paper studied these reactions employing a phenomenological resonance model finding non-negligible contributions coming from the exchange of the ground state $1/2^+$ hyperons in s- and u-channel exchange configurations.

A step further was taken in \cite{Ramos:2016odk}, where the relevance of the Born terms in the interaction kernel was tested when the $K\Xi$ channels are included in the fits. The results revealed the particular importance of the u- and s-diagrams in these channels, assigning them a role similar to the NLO contributions. This finding led to significant modifications of the NLO parameters. In such a situation one could expect a somewhat improved reproduction of the experimental data, but the set of parameters from this last fit offers very similar reproduction of the $ K^- p \to K\Xi $ scattering data to that of the best pure chiral model in \cite{Feijoo:2015yja}. The physics involved in these two parametrizations is understood after splitting the $K^- p \to K \Xi$ cross-section into the isospin basis showing a very different distribution pattern between models, whose only difference is the inclusion or not of the Born terms in the interaction kernel. This is a clear evidence of the need to explore reactions that proceed through either $I=0$ or $I=1$, thereby acting as isospin selectors from which one can extract valuable information to constrain the parameters of the meson-baryon Lagrangian.

Actually, the $K^- p \to \eta \Lambda, \eta \Sigma^0$ reactions provide us with such an opportunity since they are pure isospin $0$ and $1$ filtering processes respectively. Although experimental data \cite{Starostin,Baxter,Jones,Berthon} have already been available since the 1970s and the most recent one from the Crystal Ball Collaboration since 2001, they have barely been used in this sort of study. Therefore, motivated by the findings of \cite{Feijoo:2015yja,Ramos:2016odk},
we perform a study of the meson-baryon interaction in the $S=-1$ sector incorporating these experimental cross-section data in our approach, thus having information from all possible channels of the sector. We remark that incorporating the  $K^- p \to K^+,\Xi^-$, $K^0,\Xi^0$, $\eta\Lambda$ and $\eta\Sigma^0$ channels provides new and valuable information to constrain the NLO parameters but, at the same time, involves extending the range of energies of the model well above the $K^- p$ threshold, much beyond what is considered acceptable in usual chiral perturbation approaches. We therefore should regard the unitary chiral approach presented here as an effective chirally motivated phenomenological model that is able to describe the data on the ${\bar K}N$ interaction and related channels in a wide energy range, without compromising the good description of the low-energy data, and accommodate the new data that might become available at the experimental facilities. In this spirit,
we also give predictions for other reactions that will provide additional information in this sector, such as the $K^0_L p \to K^+ \Xi^0$ reaction, which is an $I=1$ filtering process and has been proposed to be measured at JLAB \cite{Albrow:2016ibs}, and the weak $\Lambda_b$ decay into a $J/\Psi$ and a meson-baryon pair, a reaction that filters the $I=0$ component in the final meson-baryon state \cite{Aaij:2015tga}. In the present work, we focus on the decay of the $\Lambda_b$ into $\eta \Lambda$ and $K \Xi$ final meson-baryon pairs. In addition, a study of the pole content of the model paying special attention to the poles of $\Lambda(1405)$  is included for completeness.

In \cite{Feijoo:2015yja} the inclusion of additional high-spin and high-mass resonant contributions into the $K \Xi$ scattering amplitudes plays a double role: on the one hand, it improves the description of the experimental $ K^- p \to K\Xi $ total cross sections; on the other hand, it also allows us to study the stability of the NLO coefficients because these phenomenological contributions implicitly simulate higher-angular-momenta contributions involving low lying meson-baryon states of the coupled channel problem. Otherwise, the low energy constants might absorb these contributions in order to reproduce the experimental data at the expense of taking less realistic values. Keeping this in mind, we finally perform a new fit taking into account resonant contributions where the NLO terms are considered to play a relevant role. More specifically, we consider explicit $\Lambda(1890), \Sigma(2030)$ and $\Sigma(2250)$ resonances in the $ K^- p \to K\Xi $ total cross sections, and the $\Lambda(1890)$ one in the $ K^- p \to \eta \Lambda$ process. 

The paper is organized as follows. In Sec.~\ref{theory} the theoretical aspects are presented, including a summary of the chiral unitary theory, the formalism for calculating various isospin filtering reactions, and the details on how the resonant terms are implemented. The fitting procedure and data treatment  are described in Sec.~\ref{sec:Fit_Data_treat}. The results of two different fitting models are presented and discussed in Sec.~\ref{subsec:results}, where predictions for processes that filter either the $I=0$ or the $I=1$ component of the $S=-1$ meson-baryon interactions are also given. Some final remarks and conclusions are given in Sec.~\ref{sec:conclusions}.

\section{Theoretical background}
\label{theory}

\subsection{Chiral unitary approach} 
\label{UChPT}

Since the meson-baryon interaction from effective chiral Lagrangians has been widely used and one can find plenty of literature devoted to its derivation,  we just provide the main points as a guideline and stress the particularities of our model in this section. The $SU(3)$ chiral effective Lagrangian up to NLO is taken as departing point, 
\begin{equation}
\Lagr_{\phi B}^{eff}=\Lagr_{\phi B}^{(1)}+\Lagr_{\phi B}^{(2)}  \ ,
\end{equation}
with $\Lagr_{\phi B}^{(1)}$ and $\Lagr_{\phi B}^{(2)}$ being the most general form of the LO and NLO (s-wave) contributions to meson-baryon interaction Lagrangian, respectively, expressed as follows  
\ba 
\Lagr_{\phi B}^{(1)} & = & i \langle \bar{B} \gamma_{\mu} [D^{\mu},B] \rangle
                            - M_0 \langle \bar{B}B \rangle  
                           - \frac{1}{2} D \langle \bar{B} \gamma_{\mu} 
                             \gamma_5 \{u^{\mu},B\} \rangle \no \\
                  & &      - \frac{1}{2} F \langle \bar{B} \gamma_{\mu} 
                               \gamma_5 [u^{\mu},B] \rangle \ ,
\label{LagrphiB1} 
\ea 
\begin{eqnarray}
    \Lagr_{\phi B}^{(2)}& = & b_D \langle \bar{B} \{\chi_+,B\} \rangle
                             + b_F \langle \bar{B} [\chi_+,B] \rangle
                             + b_0 \langle \bar{B} B \rangle \langle \chi_+ \rangle \no \\ 
                     &  & + d_1 \langle \bar{B} \{u_{\mu},[u^{\mu},B]\} \rangle 
                            + d_2 \langle \bar{B} [u_{\mu},[u^{\mu},B]] \rangle    \no \\
                    &  &  + d_3 \langle \bar{B} u_{\mu} \rangle \langle u^{\mu} B \rangle
                            + d_4 \langle \bar{B} B \rangle \langle u^{\mu} u_{\mu} \rangle \ .
\label{LagrphiB2}
\end{eqnarray}
In both equations, $B$ represents a $3\times3$ unitary matrix that contains the fundamental baryon octet $(N,\Lambda,\Sigma,\Xi)$. The chiral symmetry conservation requires a more complicated prescription, $u_\mu = i u^\dagger \partial_\mu U u^\dagger$, to enter the pseudoscalar meson octet ($\pi,K,\eta$), which are compacted in the $3\times3$ unitary $\phi$  matrix, and with $U(\phi) = u^2(\phi) = \exp{\left( \sqrt{2} {\rm i} \phi/f \right)} $ as the chiral fields where $f$ is the meson decay constant. The symbol $\langle \dots \rangle$ stands for the trace in flavor space. In eq.~(\ref{LagrphiB1}), $M_0$ is the common baryon octet mass in the chiral limit, and the $SU(3)$ axial vector constants $D$ and $F$ are subject to the constraint $g_A=D+F=1.26$ from the determination of hyperon and neutron $\beta$ decays \cite{Ratcliffe:1998su}. The local character of the chiral transformation of $u$ makes mandatory the introduction of a covariant derivative,  $[D_\mu, B] = \partial_\mu B + [ \Gamma_\mu, B]$ with $\Gamma_\mu =  [ u^\dagger,  \partial_\mu u] /2$ being the chiral connection, which  transforms in the same way as the baryon fields. Furthermore, in Eq.~(\ref{LagrphiB2}), we find $\chi_+ = 2 B_0 (u^\dagger \mathcal{M} u^\dagger + u \mathcal{M} u)$ which breaks chiral symmetry explicitly via the quark mass matrix  $\mathcal{M} = {\rm diag}(m_u, m_d, m_s)$ and $B_0 = - \bra{0} \bar{q} q \ket{0} / f^2$ which relates to the order parameter of spontaneously broken chiral symmetry. The coefficients $b_D$, $b_F$, $b_0$ and $d_i$ $(i=1,\dots,4)$ are the corresponding low energy constants at NLO. In principle, these constants are not fixed by the symmetries of the underlying theory, but need to be determined from experiment. Actually, the parameters accompanying the first two terms of Eq.~(\ref{LagrphiB2}), involved in terms proportional to the $\chi_+$ field,  should fulfill constraints related to the mass splitting of baryons. Once these parameters are determined, the $b_0$-coefficient could well be extracted from the pion-nucleon sigma term or from the strangeness content of the proton \cite{Gasser:1990ce}. The rest of the low energy constants, namely $d_i$, can be constrained using data coming from the meson-baryon octet such as the isospin even $\pi N$ s-wave scattering length \cite{Koch:1985bn} and the isospin zero kaon-nucleon s-wave scattering length \cite{Dover:1982zh}. Nevertheless, given that our study goes beyond tree level because of the implementation of the coupled-channel unitarization, we will release these constraints and consider the $b$-type constants, together with the $d_i$ ones, as free parameters in the fitting procedure, as usually done in the literature.

\begin{figure}[ht]
\begin{center}
\centering
\includegraphics[width=3.4 in]{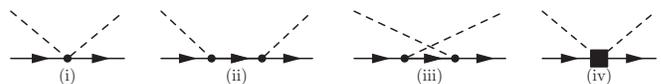}
\caption{\label{fig:epsart} Feynman diagrams for the meson-baryon interaction: Weinberg-Tomozawa term (i), direct and crossed Born terms (ii) and (iii), and NLO terms (iv). Dashed (solid) lines represent the pseudoscalar octet mesons (octet baryons).}
\end{center}
\end{figure}
At this point, the meson-baryon interaction kernel in momentum space can be derived from Eqs.(\ref{LagrphiB1}) and (\ref{LagrphiB2}). Being more precise, the WT contribution corresponds to the contact diagram (i) in Fig.~\ref{fig:epsart}; this comes from the term with the covariant derivative in Eq.~(\ref{LagrphiB1}). Next, the vertices of diagrams (ii) and (iii), which stand for the direct and crossed Born contributions, are obtained from the second and third terms of  Eq.~(\ref{LagrphiB1}); while the NLO contact one is directly extracted from (\ref{LagrphiB2}) whose representation can be found in the fourth diagram of  Fig.~\ref{fig:epsart}. This way, the total interaction kernel up to NLO is expressed as the sum:
\begin{equation}
\label{V_TOT}
 \hat{V}_{ij}=V^{\scriptscriptstyle WT}_{ij}+V^{D}_{ij}+V^{C}_{ij}+V^{\scriptscriptstyle NLO}_{ij} ,
\end{equation}
where the elements of the interaction matrix, written as $\hat{V}_{ij}= \langle i | \hat{V} | j \rangle $, couple the meson-baryon channels. The indices $(i,j)$ cover all the initial and final channels, which, in the case of the sector explored here, amount to ten: $K^-p$, $\bar{K}^0 n$, $\pi^0\Lambda$, $\pi^0\Sigma^0$, $\pi^-\Sigma^+$, $\pi^+\Sigma^-$, $\eta\Lambda$, $\eta\Sigma^0$, $K^+\Xi^-$ and $K^0\Xi^0$. The interaction $\hat{V}_{ij}$ depends on the total energy of the meson-baryon system in the center-of-mass frame $\sqrt{s}$, on the solid angle of the scattering $\Omega$, and on the $\sigma_{i},\sigma_{j}$ spin degrees of freedom of the baryons in channels $(i,j)$. Considering the scattering of a Nambu-Goldstone (NG) boson with a spin $\frac{1}{2}$ baryon target leads to a  contribution which depends only on  $\sqrt{s}$ coming from the projection of $\hat{V}_{ij}(\sqrt{s},\Omega, \sigma_{i},\sigma_{j})$ onto the s-wave:
\begin{equation}
\label{s_dependent_V}
V_{ij}(\sqrt{s})=\frac{1}{8 \pi}\sum_{\sigma}\int d\Omega \; \hat{V}_{ij}(\sqrt{s},\Omega, \sigma_{i},\sigma_{j}) . 
\end{equation}
The reader is referred to \cite{Feijoo:2015yja,Ramos:2016odk,Feijoo_thesis} for a detailed derivation of the different contributions to the final interaction kernel and their corresponding algebraic expressions.

The extension from $SU(2)$ to $SU(3)$ of the chiral Lagrangian carries an inherent effect associated with explicit chiral symmetry breaking by the strange-quark mass. Moreover, the $\bar{K}N$ interaction is strong enough to dynamically generate the $\Lambda(1405)$ resonance. Under such circumstances, a plain chiral perturbation theory (ChPT) expansion is not appropriate, making the use of a nonperturbative method absolutely necessary. The UChPT method consists of solving the Bethe-Salpetter (BS) equations, which come in a complex system of coupled-integral equations for the scattering amplitudes, once we get a well-defined potential, i.e. Eq.~(\ref{s_dependent_V}). In order to obtain the scattering amplitude, we proceed following \cite{OR,HJ_rev} where the interaction kernel is conveniently split  into its on-shell contribution and the corresponding off-shell one. The off-shell part gives rise to a tadpole-type diagram, which can be reabsorbed into renormalization of couplings and masses and could, hence, be omitted from the calculation. This procedure permits factorizing the interaction kernel and the scattering amplitude out of the integral equation, leaving a simple system of algebraic equations to be solved which, in matrix form, reads:
\begin{equation}
T_{ij} ={(1-V_{il}G_l)}^{-1}V_{lj} ,
 \label{T_algebraic}
\end{equation}
where $V_{ij}$ and $T_{ij}$ are the interaction kernel and its corresponding scattering amplitude, respectively, for a given starting i-channel and an outgoing j-channel, and $G_l$ is the loop function standing for a diagonal matrix with elements: 
\begin{equation} \label{Loop_integral}
G_l={\rm i}\int \frac{d^4q_l}{{(2\pi)}^4}\frac{2M_l}{{(P-q_l)}^2-M_l^2+{\rm i}\epsilon}\frac{1}{q_l^2-m_l^2+{\rm i}\epsilon} ,
\end{equation} 
where $M_l$ and $m_l$ are the baryon and meson masses of the $l$-channel. Since this function diverges logarithmically, the dimensional regularization is applied obtaining as final expression:
\ba
& G_l = &\frac{2M_l}{(4\pi)^2} \Bigg \lbrace a_l(\mu)+\ln\frac{M_l^2}{\mu^2}+\frac{m_l^2-M_l^2+s}{2s}\ln\frac{m_l^2}{M_l^2}  \no \\ 
 &     & + \frac{q_{\rm cm}}{\sqrt{s}}\ln\left[\frac{(s+2\sqrt{s}q_{\rm cm})^2-(M_l^2-m_l^2)^2}{(s-2\sqrt{s}q_{\rm cm})^2-(M_l^2-m_l^2)^2}\right]\Bigg \rbrace .  
 \label{dim_reg}    
\ea 
The loop function $G_l$ comes in terms of the subtraction constants $a_l$ that replace the divergence for a given dimensional regularization scale $\mu$ which is taken to be $1$~GeV in the present work. These constants are unknown, a fact that implies the inclusion of them in fitting procedures to determine their values. The number of independent subtraction constants is commonly reduced by isospin symmetry arguments and, in this particular sector, amount to 6. Nonetheless, the lack of knowledge about the values of the $a_l$ constants does not rule out the possibility of establishing a natural size for them. Indeed, as discussed in \cite{OM}, a direct comparison between the dimensional regularization method and an approximation to calculate the loop function using a cut off, which was carried out in \cite{OR}, provide the following relation (it should be noted that we use a different remapping than that of \cite{OM}):
\begin{equation}
\label{sc_nat_size}
a_l(\mu)=\frac{1}{16 \pi^2 } \Big[ 1-2 \ln \left(1+\sqrt{1+\frac{\bar{M}^2_l}{\mu^2}} \right)+\dots \Big] ,
\end{equation}
where $\bar{M}_l$ stands for the average mass of the octet of $J^\pi=\frac{1}{2}^+$, and the ellipses indicates higher order terms in the non-relativistic expansion as well as powers of $m_l/M_l$.  As we use $\mu=1$~GeV,  and taking into account these corrections, natural-sized values for the subtraction constants would lie in the range $(-10^{-2},+10^{-2})$.

The dynamically generated resonance states show up as pole singularities ($z_p=M_R-{\rm i}\Gamma_R/2$) of the scattering amplitude in the second Riemann sheet (RS) of the complex energy plane, whose real and imaginary parts correspond to its mass ($M_R$) and half width ($\Gamma_R/2$). Since the loop function is written in terms of the relative momentum of the two-body system, one is able to determine the RS of the amplitude. Practically speaking, it means that one must perform a calculation of the loop function given by Eq.~(\ref{dim_reg}) taking into account a reflection on momentum ($q_l\to -q_l$), which is equivalent to the following rearrangement on the loop function
\begin{equation}
 G_{l}^\text{II}(\sqrt{s})=G_{l}(\sqrt{s})+i\,2M_l\frac{q_l}{4\pi\sqrt{s}},
\end{equation}
for a general complex value of $\sqrt{s}$, where the superscript \text{II} denotes the rotation to the second RS. It should be stressed that, when it comes to a multi-channel sector, each channel's loop function will only be rotated to the second RS if the real part of the complex energy is larger than the corresponding channel threshold.

Having located the pole and assuming a Breit-Wigner structure for the scattering amplitude in the proximity of the found pole on the real axis,
\begin{equation}\label{eq:pole}
T_{ij}(\sqrt{s})\sim \frac{g_i g_j} {\sqrt{s}-z_p} \ ,
\end{equation}
the complex coupling strengths ($g_i$, $g_j$) of the resonance to the corresponding meson-baryon channels can be connected to the residue of the pole. 

In Ref.~\cite{Bruns:2010sv} it was discussed that some conceptual and practical drawbacks appear when the Born contributions are included in the driving term of the BS equation. Thus, the on-shell scheme described above must be treated with care. This is of special relevance for the u-channel Born diagram where the propagator of its intermediate baryon could generate non-physical sub-threshold cuts. A possible consequence is the contribution of the cuts of some heavy meson-baryon pairs to physical processes involving light meson-baryon channels. The authors of \cite{BNW} suggested to assign a constant value to the u-channel interaction kernel below a certain invariant energy in order to deal with this artifact. Such a case was detected when exploring the subthreshold analytical extension of the kernel for the $\eta \Sigma$ transition, giving rise to a cusp in the amplitude around $1423$~MeV. We proceed in the way suggested in \cite{BNW} and fix the elastic $\eta\Sigma$ u-diagram contribution to the potential to $V^{C}(\sqrt{s})=V^{C}(\sqrt{s}=1430) $ for energies $\sqrt{s}\leq1430$~MeV. This choice does not affect either the results of the observables employed here because all of them are calculated above this energy value or the $\Lambda(1405)$ properties given its $I=0$ nature. 

\subsection{Isospin filtering processes} 
\label{Isos_filters1}

As previously noted in the Introduction, the inclusion of the Born terms in the interaction kernel in \cite{Ramos:2016odk} did not improve significantly the description of data over the best pure chiral model of \cite{Feijoo:2015yja}, a fact that was already previously found \cite{BNW,Mizutani:2012gy,IHW}. Actually, in \cite{BNW,Mizutani:2012gy}, the authors obtain very similar parametrizations when studying the impact of a systematic inclusion of the Born terms into models that are based on an interaction kernel which considers WT and NLO contributions and which are fitted to the classical channels. The only eye-catching feature of the models that incorporate the u- and  s-diagrams in the interaction kernel is the more natural size of the subtraction constants. The same finding applies, of course, to the model developed in  \cite{Ramos:2016odk}. But, in contrast to other groups, the effect of including the $K\Xi$ scattering data in the fits led to a very different set of NLO parameters than those obtained from the best chiral model of \cite{Feijoo:2015yja}, which neglects the Born terms and was fitted to the same experimental data. These differences denote the significant role played by the Born contributions once the $ K \Xi$ channels are taken into account. 
A detailed look at the total cross sections obtained from the models of \cite{Feijoo:2015yja,Ramos:2016odk} makes one realize that there is not any substantial difference between the models when reproducing the classical channels. The explanation stems from the similar $f$ values for these two models combined with the well-known dominant role played by the WT term for the former channels. This is not the case for the $K^- p\to K \Xi$ reactions.
Apart from the differences in reproducing the details of the structure shown by the experimental points, both models present opposite patterns in cross-section strength for the $K^+ \Xi^-$ and $K^0 \Xi^0$ channels despite peaking at the same energies [see Fig.~(5) in \cite{Ramos:2016odk}]. Obviously, these findings indicate that these models predict a different distribution of the isospin amplitudes. Indeed, the breakdown into the $I=0$ and $I=1$ components of the $ K^- p \to K\Xi $ total cross sections performed in Ref.~\cite{Ramos:2016odk} reflects this fact, being more evident for the $I=0$ components, where the size and distribution of strength predicted by the two models was very different. This observation points towards using data coming from isospin filtering reactions as a tool to discriminate between the models and, what is even more important, to provide more solid constraints on the fitting parameters. %

The meson-baryon interaction in the $S=-1$ sector comprises several reactions with single-isospin outgoing channels. Particularizing those that come from $K^-p$ scattering, one has the production of $\eta \Lambda$ with $I=0$ and the $\pi \Lambda, \eta \Sigma^0$ productions with $I=1$. The $\pi \Lambda$ experimental data has been widely employed by most of the previous cited models. By contrary, as far as we know, the available $K^- p \to \eta \Lambda$ scattering data has only been used in \cite{GO,Kamano:2014zba} and we were not able to find a reference related to the use of $K^- p \to \eta \Sigma^0$ data. Given this situation, the next natural step is the study of the effects on the fitting parameters when the scattering data from the $K^- p \to \eta \Lambda, \eta \Sigma^0$ reactions is included in the fits, as done in the present work. In Sec.~\ref{subsec:results}  it will be shown that these channels have a positive impact on the NLO coefficients and on the subtraction constants.
 
\subsubsection{$K_L^0p \to K^+ \Xi^0$ reaction}
\label{Isos_filters2}

The K-Long Facility for JLab has planned the measurement of two-body reactions induced by a secondary $K^0_L$ beam on a liquid hydrogen cryotarget to improve the knowledge of the $\Lambda^*$ and $\Sigma^*$ spectroscopy \cite{Albrow:2016ibs}. To this end, they propose several reactions ($K_L^0p \to K_s p, \pi^+ \Lambda, K^+ \Xi^0, K^+ n, K^- \pi^+ p$) to be explored within a center-of-mass (c.m.) energy from $1490$ to $4000$~MeV. Given that $K^0_L = (K^0 -{\bar K}^0)/\sqrt{2}$, the former reactions would proceed through the ${\bar K}^0$ component of the $K^0_L$, and, thus, would be of pure isovector character.  As our model is purely s-wave, it can study the $K_L^0p \to K^+ \Xi^0$ reaction, which will be measured sufficiently close to threshold. Nevertheless, the experimental data of this process will provide valuable constraints for NLO low-energy constants because it involves $K\Xi$ channels in the $S=-1$ and $Q=+1$ sector whose amplitudes can be related to the ones of our sector ($S=-1$, $Q=0$), employing isospin symmetry arguments.  Indeed, the amplitude for the JLab process can be written in terms of the strong-interaction states in $| I I_3\rangle=| 1 1 \rangle$,
\begin{equation}
\label{scat_ampli_Jlab_1}
\langle K^+ \Xi^0 |T | K^0_L p \rangle  =  -\frac{1}{\sqrt{2}}\langle K^+ \Xi^0 | T | \bar{K}^0 p \rangle ,
\end{equation}
and, invoking  the invariance of the strong interaction under $ I_3 $ rotations, this amplitude can be expressed in terms of states with $ I_3=0 $, which are the ones employed in our studies, as 
\begin{eqnarray}
\label{scat_ampli_Jlab_2}
& & \langle K^+ \Xi^0 |T | K^0_L p \rangle   =  -\frac{1}{2 \sqrt{2}}\big[ \langle K^0 \Xi^0 | T | \bar{K}^0 n \rangle - \langle K^0 \Xi^0 | T | K^- p \rangle   \no  \\
& & - \langle K^+ \Xi^- | T | \bar{K}^0 n \rangle + \langle K^+ \Xi^- | T |  K^- p \rangle \big] .  
\end{eqnarray}
Even though there is no experimental data for this reaction at the moment, the only two points obtained from the $K^-$ deuteron reactions on bubble chamber experiments \cite{iso1exp1,iso1exp2} can be employed as baseline for our predictions. These two experimental points coming from $K^- n \to  K^0 \Xi^-$ cross section must be divided by $2$ to properly account for the size of the $S=-1$ component of  $K^0_L p$. This could be easily seen by performing a calculation similar to the one done previously to finally find that both sections are related by
\begin{eqnarray}
\label{cross_sect_relationship}
\sigma_{K^- n \to  K^0 \Xi^-}  & \propto  &  | \langle K^0 \Xi^- |T | K^- n\rangle |^2=2 |\langle K^+ \Xi^0 |T | K^0_L p \rangle|^2  \no  \\
 \sigma_{K^0_L p \to K^+ \Xi^0}  & =  & \frac{1}{2}\sigma_{K^- n \to  K^0 \Xi^-} .
\end{eqnarray}
\subsubsection{ $\Lambda_b \to J/\Psi ~ K\Xi, ~ J/\Psi ~ \eta \Lambda$ decays}
\label{Isos_filters3}

Another opportunity to learn about the isoscalar component of the meson-baryon interaction of interest here comes from the weak decay of the $\Lambda_b$ into final states containing a $J/\Psi$ and a $S=-1$  meson-baryon pair. The pioneering theoretical study of Ref.~\cite{Roca:2015tea} focuses on the $\Lambda_b \to J/\psi ~K^- p (\pi \Sigma)$ decay, finding that this type of reaction does filter the final meson-baryon components in $I=0$. Subsequent experimental findings at LHCb \cite{Aaij:2015tga} supported this idea showing that this specific decay is dominated by intermediate $\Lambda^*$ resonances and, in particular, they confirm the contribution of the $\Lambda (1405)$ tail in the $K^- p$ invariant mass distribution predicted by the authors of \cite{Roca:2015tea}.  The success of this mechanism has triggered a lot of activity in the community, the recent study of $\Lambda_b \to \eta_c ~K^- p (\pi \Sigma)$ decay \cite{Xie:2017gwc} is an evidence of it, where the authors show that the final-state interaction is basically mediated by the $\Lambda (1405)$ and where it is highlighted that this decay is a very clean one, free of any I =1 contribution, to produce such resonance. Even more recently, the theoretical analysis in \cite{Miyahara:2018lud} provides new insights on $\pi \Sigma$ spectra in the region of the $\Lambda (1405)$ by means of exploring the meson-baryon final state interaction in the $\Xi^0_b$ weak decay into $D^0$ and a meson-baryon pair. There, it is shown that the $\pi^+ \Sigma^-$ channel is the only suitable one to provide a signal for the resonance due to the interference effect between the direct and rescattering processes. Although the production rate of $\Xi_b$ relative to that of $\Lambda_b$ has a much poorer statistics \cite{Aaij:2017bef}, ongoing and future measurements of these heavy-hadron weak decays will provide new opportunities to obtain more information about this decay mechanism and the $\Lambda (1405)$ puzzle.

Prior to the previous references yet in the same footing, in \cite{Feijoo:2015cca}, the $\Lambda_b \to J/\Psi K^+ \Xi^- (\eta \Lambda)$ reactions were studied aiming for making progress in our understanding of hadron dynamics at higher energies. A comparison between the experimental  $K \Xi$ and $\eta \Lambda$ invariant masses with the theoretical predictions would reveal information of the low-energy constants, given the sensitivity of the final meson-baryon interaction in these two processes to the NLO terms.  In this respect, we present new results for these processes when we employ the interaction model developed here, which uses new isospin filtering data in the fits as well as the Born terms in the driving potential.

\begin{figure}[!htb]
\center
 \includegraphics[width=0.9\linewidth]{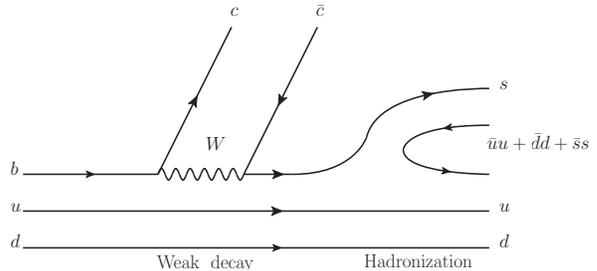}
\caption{Diagram describing the weak decay of the
$\Lambda_b$ into the $J/\psi$ and a meson-baryon pair formed
through a hadronization mechanism.}\label{fig:decay}
\end{figure}
These reactions involve an elementary weak transition at the quark level that proceeds via the creation of a $J/\psi$ meson and an excited $sud$ system with $I=0$ that hadronizes into a final $I=0$ meson-baryon pair. The Cabibbo favored mechanism for $J/\psi$ production is depicted by the first part of the diagram of Fig.~\ref{fig:decay}, where one can see the $W$-exchange weak process transforming the $b$ quark into $c{\bar c} s$. As, in this topology, the $u$ and $d$ quarks remain as spectators, they keep the $I=0$ configuration they had in the $\Lambda_b$ and, consequently, the $sud$ state after the weak decay must also be in $I=0$. This is in good agreement with the analysis of the experimental $K^-p$ invariant mass performed by the LHCb Collaboration, as we have already advanced. The next step consists of the hadronization of $sud$ state, which requires the creation of $\bar{q} q$ pairs with the quantum numbers of the vacuum, i.e.,  in the form $\bar u u+ \bar d d+ \bar s s$, which ends up producing meson-baryon states in the final state. The technical details of the hadronization can be found in \cite{Roca:2015tea,Feijoo:2015cca,Oset:2016lyh}. Here, we merely give the resulting state $|H\rangle$ in terms of its meson-baryon components
\begin{equation}
|H\rangle=|K^-p\rangle+|\bar{K}^0n\rangle+\frac{\sqrt{2}}{3}|\eta\Lambda\rangle \ ,
\label{eq:primaryprod}
\end{equation}
where the $|\eta'\Lambda\rangle$ contribution has been omitted on account of the  $\eta'$ large mass \cite{Roca:2015tea}.  

\begin{figure}[!htb]
\center
 \includegraphics[width=0.9\linewidth]{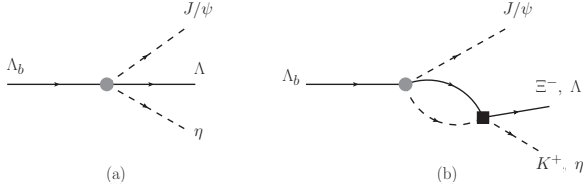}
\caption{Diagrammatic representation of the decay amplitude for $\Lambda_b \to J/\psi \, \phi_j B_j $: (a) tree level and (b) the $\phi_j M_j=\eta\Lambda , K^+ \Xi^- $ production through the coupled channel interaction of the initially produced $\phi_i M_i=\eta\Lambda , \bar{K}N$ meson-baryon pairs.}\label{fig:diagrams}
\end{figure}

Finally, the amplitudes for the $\Lambda_b$ decay into $J/\psi \, \eta \Lambda$ and  $J/\psi \, K^+ \Xi^- $ states split into two contributions: the direct tree-level process and the final-state interaction contribution of the primary meson-baryon pair into final $\eta \Lambda$ or $ K^+ \Xi^- $ production, depicted in Figs.~\ref{fig:diagrams}(a) and \ref{fig:diagrams}(b), respectively.  This amplitude can be written as:
\begin{equation}
\mathcal{M}(M_{\phi B},M_{J/\psi B}) = V_{p} \Big[ h_{\phi B} + \sum_i h_i G_i( M_{\phi B}) t_{i,\phi B}( M_{\phi B}) \Big] ,
\label{eq:amplitude}
\end{equation}
where the weights $h_i$, obtained from Eq.~(\ref{eq:primaryprod}), are:
\begin{eqnarray}
&h_{\pi^0\Sigma^0}=h_{\pi^+\Sigma^-}=h_{\pi^-\Sigma^+}=0\,,~h_{\eta\Lambda}=
\frac{\sqrt{2}}{3}\,,\\
&h_{K^-p}=h_{\bar K^0n}=1\,,~h_{K^+\Xi^-}=h_{K^0\Xi^0}=0\ ,
\end{eqnarray}
$G_i$, with $i={K^-p, \bar{K}^0 n, \eta\Lambda}$, denotes the one-meson-one-baryon loop function [see Eq.~(\ref{dim_reg})] and the amplitude $t_{i,\phi B}$ is chosen in accordance with the models employed in the present study. Here, $\phi B$ can be either $\eta\Lambda$ or $K \Xi$ and $M_{\phi B},M_{J/\psi B}$ stand for the corresponding invariant masses. As an interesting observation, it should be mentioned that the production of $K^+ \Xi^-$ states is only allowed from the rescattering of meson-baryon components and, hence, the  $\Lambda_b \to J/\psi \, K^+ \Xi^- $ decay process depends strongly on the meson-baryon interaction model employed. The factor $V_p$, which includes the common dynamics of the production of the different pairs, is unknown and we take it as constant, which means that the decay distributions will not have units. This also implies that the form factors at the weak vertex have been assumed to behave  smoothly with energy, so that the energy dependence of $\mathcal{M}(M_{\phi B},M_{J/\psi B})$ in Eq.~(\ref{eq:amplitude}) can be associated essentially to the changes of the final state interaction. There is a thorough discussion to support this point in Ref.~\cite{Feijoo:2015cca}.

The double differential cross-section for the $\Lambda_b \to J/\psi \, \phi B$ decay process reads:
\begin{eqnarray}
&&\frac{d^2\Gamma}{dM_{\phi B}dM_{J/\psi B}}  =  \frac{1}{{(2\pi)}^3}\frac{4M_{\Lambda_b}M_{B}}{32M_{\Lambda_b}^3} \nonumber \\ 
&& ~~~~~~\times \overline{\sum}| \mathcal{M}(M_{\phi B},M_{J/\psi B})|^2 2 M_{\phi B} 2 M_{J/\psi B}  \  , 
\label{eq:double_diff_cross}
\end{eqnarray}
where $ \overline{\sum}$ stands for the sum over final spins and polarizations and the average over initial spins, which can be replaced by a factor 3 (see the Appendix in \cite{Lu:2016roh}). Since we are interested in presenting the results in terms of the invariant masses $M_{\eta \Lambda}$ and $M_{K^+ \Xi^-}$, we fix the invariant mass $M_{\phi B}$ and integrate expression (\ref{eq:double_diff_cross}) over $M_{J/\psi B}$ in order to obtain $d\Gamma/dM_{\phi B}$.

\subsection{Inclusion of high spin hyperon resonances}
\label{theory_resonances}

Encouraged by the improved data description when adding resonant contributions into the $K\Xi$ cross sections obtained by chirally motivated models \cite{Feijoo:2015yja}, we extend this procedure to the $K^- p \to \eta \Lambda$ reaction in the present work. In addition to this phenomenological reason, the  incorporation of resonances also offers the possibility of checking the stability of the NLO coefficients. The resonant terms take into account, in an effective way, higher angular momentum contributions and, in principle, the more relevant the higher-angular-momenta terms are, the further the low energy constants will be from their nominal values in the absence of such contributions. Therefore, the resonant contributions permit these parameters to get relaxed, avoiding a possible overestimation of their values. 

Currently, eight three- and four-star status resonances with masses lying in the range $1.89<M<2.35$ GeV are listed in the PDG compilation \cite{PDG}. However, guided by \cite{Jackson:2015dva,Feijoo:2015yja,Sharov:2011xq} and after an exhaustive inspection of the effects of all the possible resonances on the  $ K^- p \to K\Xi, \eta\Lambda $ cross sections, we conclude that the candidates which reproduce the experimental data better are $\Sigma(2030)$,  $\Sigma(2250)$ and $\Lambda(1890)$ with spin-parity $J^\pi=7/2^+$, $5/2^-$ and $3/2^+$, respectively. It should be clarified that, although the $J^\pi$ quantum numbers for $\Sigma(2250)$ are not known, we choose $J^\pi =5/2^-$, which is one of the most probable assignments. If the $9/2^-$ choice had been made, the calculation would have been unnecessarily complicated since it would not have changed the results drastically \cite{Sharov:2011xq}.
\begin{figure}[!htb]
\center
 \includegraphics[width=0.65\linewidth]{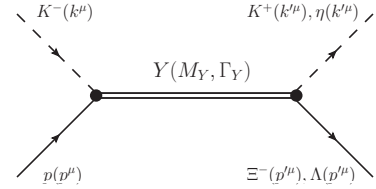}
\caption{S-type diagram describing the resonant amplitude, where the intermediate hyperon ($Y$= $\Lambda(1890)$, $\Sigma(2030)$, $\Sigma(2250)$) has mass $M_{J}$ and decay width $\Gamma_{J}$.}\label{fig:resonant_diagram}
\end{figure}

The high-spin resonant contributions require a special treatment, analogous to that performed in \cite{Nakayama:2006ty,Man:2011np,Sharov:2011xq}. This consists of adopting the Rarita-Schwinger method that permits building resonant amplitudes from effective Lagrangians, which describe spin-$3/2$, -$5/2$ and -$7/2$ baryon fields as rank-$1$, -$2$ and -$3$ tensors ($Y_{3/2}^{\mu}, ~ Y_{5/2}^{\mu\nu}$ and $Y_{7/2}^{\mu\nu\alpha}$),respectively,
\ba
\mathcal{L}^{{7/2}^\pm}_{BY\phi}(q)&=&-\frac{g_{BY_{7/2}\phi}}{m_K^3}\bar{B}\Gamma^{(\mp)}Y_{7/2}^{\mu\nu\alpha}\partial_\mu\partial_\nu \partial_\alpha K+H.c , \no \\
\mathcal{L}^{{5/2}^\pm}_{BY\phi}(q)&=& {\rm i}\frac{g_{BY_{5/2}\phi}}{m_K^2}\bar{B}\Gamma^{(\pm)}Y_{5/2}^{\mu\nu}\partial_\mu\partial_\nu K+H.c. , \no \\
\mathcal{L}^{{3/2}^\pm}_{BY\phi}(q)&=& {\rm i}\frac{g_{BY_{3/2}\phi}}{m_K}\bar{B}\Gamma^{(\pm)}Y_{3/2}^{\mu}\partial_\mu K+H.c. \ ,
\label{L_ResJ}
\ea
where $\Gamma^{(\pm)}= \displaystyle\binom{\gamma_5}{1}$ depending on the parity of the resonance being studied, and $g_{BY_J \phi}$ stands for the baryon-meson-$Y_{J}$ coupling. 

The implementation of the $\bar{K} N \rightarrow Y \rightarrow K \Xi (\eta \Lambda)$ transition amplitudes has been carried out by the standard s-channel diagram which is represented in Fig.~\ref{fig:resonant_diagram}. This is the simplest way to systematically approach the study of the parameter stability, which is in agreement with the treatment above the $K\Xi$ threshold developed in \cite{Sharov:2011xq}. The vertices present in the diagram can be derived from Eqs.~(\ref{L_ResJ}), and their explicit analytical form as well as the corresponding propagators, which depend on the spin and parity of the intermediate resonance, can be found in \cite{Feijoo:2015yja, Feijoo_thesis}. The resonant contributions to the scattering amplitudes can then be obtained straightforwardly as:
\ba 
& & T^{{7/2}^+}_{s',s}=F_{7/2}(k,k') \, \bar{u}_B^{s'}(p')k'_{\alpha}k'_{\beta}k'_{\sigma}S_{7/2}(q) k^{\delta}k^{\mu}k^{\nu} u_N^{s}(p) , \no \\ 
& &T^{{5/2}^-}_{s',s} = F_{5/2}(k,k')\, \bar{u}_B^{s'}(p')k'_{\alpha}k'_{\beta}S_{5/2}(q)k^{\delta}k^{\mu}u_N^{s}(p) ,   \no \\
& &T^{{3/2}^+}_{s',s} = F_{3/2}(k,k') \, \bar{u}_B^{s'}(p') \gamma_5 k'_{\alpha}S_{3/2}(q)k^{\delta} \gamma_5 u_N^{s}(p)  , 
 \label{T_J}
\ea
where $u_X^{s}$ is the spinor structure of a baryon with spin $s$, while the propagator is incorporated by means of the tensorial structures $S_{J}(q)=S_{\alpha_1\dots \alpha_{2J-1}}^{\beta_1\dots \beta_{2J-1}}(q)$ and $F_J(k,k')$ contains the couplings of the resonance to the meson-baryon channels and a Gaussian-type form factor that suppresses high powers of the meson momentum:
\begin{equation} 
F_J(k,k')=\frac{g_{B Y_{J}\phi\phantom{\bar{K}}} \!\!\! g_{NY_{J}\bar{K}}}{m_K^{2J-1}}e^{-\vec{k}^2/\Lambda^2_{J}}
e^{-\vec{k'}^2/{\Lambda'}^2_{J}}  .
\label{FF}
\end{equation} 
We have adopted this exponential prescription from the resonance based model of \cite{Sharov:2011xq}. As pointed out in our earlier work \cite{Feijoo:2015yja}, strictly speaking, these exponential factors are not genuine form factors, since these should depend on the off-shell momentum of the off-shell particle and should be normalized to $1$ at the on-shell point. They should be regarded as {\it ad hoc}  functions introduced to tame the high energy behavior  of the resonant contributions. 

Summarizing, the resonant contributions are only taken into account in the $K^- p \rightarrow Y \rightarrow K \Xi$ and $K^- p \rightarrow X \rightarrow \eta \Lambda $ transition amplitudes, where $Y$ stands for $\Lambda (1890)$, $\Sigma(2030)$ and $\Sigma(2250)$, while $X$ stands for $\Lambda (1890)$ . The scattering amplitudes are then rewritten according to the following prescription:
 \begin{itemize}
\item  For $K^- p \to K^0 \Xi^0, K^+ \Xi^-$ processes, we have
\begin{equation}
T_{ij}(s',s) = T^{BS}_{ij}(s',s)\,+\frac{1}{\sqrt{4M_pM_\Xi}}\sum_{J^\pi}T_{ij}^{{J^\pi}}(s',s) \ ,
\label{T_reso_KXi}
\end{equation}
\item and, for $K^- p \to \eta \Lambda$, 
\begin{equation}
T_{ij}(s',s) = T^{BS}_{ij}(s',s)\,+\frac{1}{\sqrt{4M_pM_\Lambda}} T_{ij}^{{3/2}^+}(s',s) \ ,
\label{T_reso_etaL}
\end{equation}
\end{itemize}
where $T^{BS}_{ij}$ is the scattering amplitude obtained from the chiral Lagrangian and unitarized by means of the BS equation, while $T^{J^\pi}_{ij}$ accounts for the corresponding resonant term with $J^\pi$ quantum numbers, which take the values $J^\pi=3/2^+,5/2^-,7/2^+$ in Eq.~(\ref{T_reso_KXi}). We note that the former contributions to the scattering amplitudes contain the appropriate Clebsh-Gordan coefficients projecting the $i$ and $j$ states into the isospin $0$ or $1$ of the resonance. The prefactor $1/\sqrt{4M_iM_j}$ has been included to properly normalize the resonant contributions in accordance with \cite{Feijoo:2015yja}. Let us point out that we could have inserted the resonant terms with a real bare mass in a unitarized scheme, the effect of which would have been to displace the mass of the resonance and to provide it with a width. Employing the physical mass and width in a non-unitarized resonant amplitude, as done here, is a simpler approach and gives effectively similar results as employing a bare real mass in a unitarizable resonant term.

\section{Fitting procedure and Data treatment}
\label{sec:Fit_Data_treat}

In the context of UChPT, the $S=-1$ sector offers us a good chance to extract information about the parameters that are present in the model from the ChPT expansion. Our model is derived from a chiral Lagrangian up to NLO in s-wave, which involves a number of low-energy constants that could amount to a maximum of 16: the meson decay constant $f$, the axial vector couplings $D$ and $F$, the NLO coefficients $b_0$, $b_D$, $b_F$, $d_1$, $d_2$, $d_3$, $d_4$; and six subtraction constants $a_{\pi \Sigma}$, $a_{\bar{K} N}$, $a_{\pi \Lambda}$, $a_{\eta \Sigma}$, $a_{\eta \Lambda}$, $a_{K \Xi}$. 

The $f$ parameter has systematically taken larger values than the experimental one in the models based on UChPT, with values ranging from $f=1.15 f_{\pi}$ to $f=1.36 f_{\pi}$, meaning to be a sort of average over the decay constants of the mesons involved in the various coupled channels. Actually, some authors assigned fixed values to this parameter \cite{GO,Mai:2014xna} depending on the meson involved in the incoming and outgoing channels ($f_{\pi}=92.4$~MeV and $f_{K}=110.0$~MeV \cite{PDG}, $f_{\eta}=120.1$~MeV estimated from the results in \cite{Kaiser:1998ds}) for each given meson-baryon process in the $S=-1$ sector. Other authors \cite{IHW} allow these three values to vary slightly or, as in our case, there are authors \cite{GO,Mizutani:2012gy} that prefer to employ a single effective $f$ parameter which should be fitted to the experimental data. 
With regard to the axial vector couplings $D$ and $F$, some works assign them fixed values, ($D=0.8$, $F=0.46$)  \cite{IHW,GO}, hence fulfilling the constraint $g_A=F+D=1.26$. We allow them to vary within 12.5\% of their canonical value in order to accommodate to the dispersion of values seen in the literature.
For the NLO parameters, the situation is also diverse. Although with some exceptions \cite{KSW,KWW,GO},  most of the performed fits have relaxed the constraints over the $b_i$ parameters, related to the mass splitting of baryons and the pion-nucleon sigma term as discussed after Eq.~(\ref{LagrphiB2}), due to the unitary character of the amplitudes. This is also the approach followed in the present work. 

Having reported on the role and importance of the different parameters present in our chiral model, we perform a fit called WT+Born+NLO,  which determines an effective set of low-energy constants that is valid in a wide energy range, including the description of the $K^-p \to K^+\Xi^-$, $K^0 \Xi^0$, $\eta \Lambda$ and $\eta \Sigma$ reactions. This fit corresponds to a unitarized calculation employing the chiral Lagrangian up to NLO, that is, an interaction kernel which incorporates the contribution of the WT, the Born and the NLO terms. 

All the observables employed in the fit require the knowledge of the $T$ matrix, which is given by Eq.~(\ref{T_algebraic}) in Sec.~\ref{UChPT}. The unpolarized total cross-section for the $i \rightarrow j$ reaction can be defined according to our normalization as:
\begin{equation}
\sigma_{ij}=\frac{1}{4\pi}\frac{M_iM_j}{s}\frac{k_j}{k_i}S_{ij} \ ,
 \label{sigma}
\end{equation}
where $s$ is the square of the c.m. energy, and where we have averaged over the initial baryon spin projections and resumed over the final ones:
\begin{equation}
\quad S_{ij}=\frac{1}{2}\sum_{s',s}|T_{ij}(s',s)|^2 \ .
 \label{M_matrix}
\end{equation}
We considered a large amount of cross section data for $K^-p$ scattering into different final channels \cite{exp_1,exp_2, exp_3, exp_4, exp_5, exp_6, exp_7, exp_8, exp_9, exp_10, exp_11,Starostin,Baxter,Jones,Berthon}. with the exception, for consistency, of the same three points that were disregarded in the fits of \cite{Ramos:2016odk} due to their strong deviation from the main trend. This fact leaves us with a total of 219 experimental points coming from $K^-p$ scattering which are collected in Table~\ref{tab_exp_points}. 

We also fit the parameters of our model to the measured branching ratios of cross section yields \cite{br_1,br_2}. These can be obtained from the elastic and inelastic $K^-p$ cross sections, Eq.~(\ref{sigma}), evaluated at threshold: 
\begin{eqnarray}
& & \gamma  =  \frac{\Gamma(K^- p \rightarrow \pi^+ \Sigma^-)}{\Gamma(K^- p \rightarrow \pi^- \Sigma^+)}= 2.36 \pm 0.04 \no ,\\ 
& & R_c = \frac{\Gamma(K^- p \rightarrow \pi^+ \Sigma^-,\pi^- \Sigma^+ )}{\Gamma(K^- p \rightarrow {\rm inelastic\, channels})}=0.189 \pm 0.015 \no , \\
& & R_n = \frac{\Gamma(K^- p \rightarrow \pi^0 \Lambda)}{\Gamma(K^- p \rightarrow {\rm neutral\, states})} = 0.664 \pm 0.011 .  
\label{branch_ratios} 
\end{eqnarray}  
\begin{table}
\centering
\begin{tabular}{lclc}
\hline \\[-2.5mm]
 {Observable}\, & \, {Points} \, & \, {Observable}\, & \, {Points} \\
\hline \\[-2.5mm]
$\sigma_{K^-p \to K^-p}$ & 23 & $\sigma_{K^-p \to \bar{K}^0n}$ & 9 \\
$\sigma_{K^-p \to \pi^0\Lambda}$ & 3 & $\sigma_{K^-p \to \pi^0\Sigma^0}$ & 3 \\
$\sigma_{K^-p \to \pi^-\Sigma^+}$ & 20 & $\sigma_{K^-p \to \pi^+\Sigma^-}$ & 28 \\
$\sigma_{K^-p \to \eta\Sigma^0}$ & 9 & $\sigma_{K^-p \to \eta\Lambda}$ & 49 \\
$\sigma_{K^-p \to K^+\Xi^-}$ & 46 & $\sigma_{K^-p \to K^0\Xi^0}$ & 29 \\
$\gamma$ & 1 & $\Delta E_{1s}$ & 1 \\
$R_n$ & 1 & $\Gamma_{1s}$ & 1 \\
$R_c$ & 1 &  & \\
\hline
\end{tabular}
\caption{Number of experimental points used in our fits,  which are extracted from \cite{exp_1,exp_2, exp_3, exp_4, exp_5, exp_6, exp_7, exp_8, exp_9, exp_10, exp_11,br_1,br_2,SIDD,Starostin,Baxter,Jones,Berthon}, distributed per observable.}
\label{tab_exp_points}
\end{table}

The $K^- p$ scattering length is obtained from the $K^- p$ scattering amplitude at threshold as:
\begin{equation}
a_{\scriptscriptstyle K^- p}=-\frac{1}{4\pi}\frac{M_p}{M_p+M_{\bar K}}T_{K^- p \to K^- p} .
 \label{scat_lenght}
\end{equation}
This scattering length is related to the energy shift and width of the $1s$ state of kaonic hydrogen via the second order corrected Deser-type formula \cite{Meissner:2004jr}:
\begin{equation}
\Delta E-i\frac{\Gamma}{2}=-2\alpha^3\mu_{r}^{2} a_{\scriptscriptstyle K^- p} \Big[ 1+2 a_{\scriptscriptstyle K^- p}\,\alpha\,\mu_r\, (1-\ln\alpha) \Big],                                                                                                                       
\label{ener_shift_width}
 \end{equation} 
where $\alpha$ is the fine-structure constant and $\mu_r$ is the reduced mass of the $K^-p$ system. These important empirical constraints were recently measured by the SIDDHARTA Collaboration \cite{SIDD}: $\Delta E_{1s}=283 \pm 36(stat) \pm 6(syst)$ eV and $\Gamma=541 \pm 89(stat) \pm 22(syst)$ eV. The distribution of points per observable is summarized in Table~\ref{tab_exp_points}.

The fitting procedure followed here consists of minimizing the $\chi^2$ per degree of freedom ($\chi^2_{\rm d.o.f.}$). The standard definition of the $\chi^2_{\rm d.o.f.}$ could skew the outputs of the fits, particularly, when the observables present substantial differences in the number of experimental points within their sets. Hence, this definition could favor the observables with a larger number of experimental points over those with a smaller amount. This bias is avoided, for instance, by adopting the method already used in  \cite{GarciaRecio:2002td,GO,IHW,Mai:2014xna}, which takes a normalized $\chi^2_{\rm d.o.f.}$ that assigns equal weight to the different measurements, namely
\begin{equation}
\chi^{2}_{\rm d.o.f}=\frac{\sum_{k=1}^K n_k }{\left( \sum_{k=1}^K n_k -p \right)} \frac{1}{K} \sum_{k=1}^K \frac{\chi^{2}_{k}}{n_k}
\label{Chi^2_dof}
\end{equation}
with
\begin{equation}
 \chi^{2}_{k}=\sum_{i=1}^{n_k}\frac{\left( y_{i,k}^{\rm th}- y_{i,k}^{\rm exp}\right)^2}{\sigma_{i,k}^{2}} \ .\nonumber 
\end{equation}
In these expressions $y_{i,k}^{\rm exp}$, $y_{i,k}^{\rm th}$ and $\sigma_{i,k}$ represent, respectively, the experimental value, theoretical prediction and error of the $i^{th}$ point of the $k^{th}$ observable, which has a total of $n_{k}$ points, $K$ is the total number of observables, and $p$ denotes the number of free fitting parameters. As it can be appreciated from Eq.~(\ref{Chi^2_dof}), the renormalizing effect is achieved by averaging the $\chi^2$ per degree of freedom over the different experiments.  The analysis for the $\chi^2$ function was carried out by means of the minimization techniques embedded in the MINUIT package. 

Error bands are also estimated for the $K^-p$ scattering cross sections into different final meson-baryon channels for the previous fit. The method followed is based on that proposed in \cite{Lampton:1976an}, and more recently employed by the authors of Ref. \cite{GO} for the definition of $\chi^2$ given by Eq.~(\ref{Chi^2_dof}). First of all, we generate new parametrizations that consist of a random variation of all the free parameters around their central values, within a wide enough band, and only keep those configurations for which the  $\chi^2$ value (Eq.~(\ref{Chi^2_dof})) satisfies
\begin{equation}
\chi^2 \leq \chi_{0}^2+\chi^2(p,1\sigma)\ ,
\label{chi_cond}
\end{equation}
where $\chi_{0}^2$ corresponds to the minimum found by MINUIT and $\chi^2(p,1\sigma)$ is the value of a chi-squared distribution with a number $p$ of degrees of freedom at a confidence level of $1\sigma$. Then, within the good configurations, we look for the maximum and minimum of each parameter, values that are then employed to determine its corresponding correlated error band.

As previously mentioned, we also perform another fit, WT+NLO+Born+RES, with the aim of  studying the effects of the resonant terms on the low energy constants. The model contains the same chiral part together with the high spin resonance contributions in the $K^- p \to K^+ \Xi^-$, $K^0 \Xi^0$, $\eta \Lambda$ channels in the way specified in Sec.~\ref{theory_resonances}. Concerning the fitting technicalities, the part coming from the chiral model involves the fitting of the same 16 parameters as in the former model.  With respect to the resonant part, we add 13 new parameters, namely: masses and widths of the resonances ($M_{Y_{3/2}}$, $M_{Y_{5/2}}$, $M_{Y_{7/2}}$, $\Gamma_{3/2}$, $\Gamma_{5/2}$ and $\Gamma_{7/2}$), which are present in the resonance propagator \cite{Feijoo:2015yja}, the product of couplings  ($g_{\Lambda Y_{3/2} \eta} \cdot  g_{NY_{3/2} \bar{K}}$, $g_{ \Xi Y_{3/2} K} \cdot g_{NY_{3/2}\bar{K}}$, $ g_{ \Xi Y_{5/2} K} \cdot g_{NY_{5/2} \bar{K}}$ and $g_{ \Xi Y_{7/2}K} \cdot g_{NY_{7/2}\bar{K}}$) and the cut-off in the form factors ($\Lambda_{3/2}$, $\Lambda_{5/2}$ and $\Lambda_{7/2}$). The number of fitting parameters amounts to a total of 29, but we would like to remark that not all parameters are fully free. The masses and widths of the resonances are taken to lie within the ranges given in the PDG compilation \cite{PDG} and the form-factor cut-off values are constrained in the range $500$~MeV$<\Lambda_{J}<1000$~MeV. This fit considers the same amount of experimental data as the previous one (Table~\ref{tab_exp_points}).

\section{Results and discussion}
\label{subsec:results}

This section is devoted to present the results obtained with the two models described in the previous section, with interaction kernels based on the WT+Born+NLO contributions, with or without the resonant terms, and both fitted to the same amount of experimental data summarized in Table~\ref{tab_exp_points}. We focus first on the outputs provided by the WT+Born+NLO fit and, subsequently, we discuss the results for the observables involved in the fitting procedure as well as the analysis of the spectroscopy from its pole content. To conclude, we show our predictions for isospin filtering processes, such as the $K^0_L p \to K^+ \Xi^0$ reaction and  the $\Lambda_b \to J/\Psi ~ K\Xi, ~ J/\Psi ~ \eta \Lambda$ decays. At the end of this section, we proceed analogously for the WT+Born+NLO+RES model but paying special attention to the NLO parameters.

The parameters of the WT+Born+NLO fit are displayed in the first column of Table~\ref{tab:outputs_fits}. Compared to the values obtained in our previous studies \cite{Feijoo:2015yja,Ramos:2016odk}, there is an overall improvement precision. Another remarkable result is that we get natural-sized values for all the subtraction constants, while, in \cite{Ramos:2016odk}, the value corresponding to $a_{\pi\Lambda}$ took an unexpectedly large value forced by the fit to accommodate the experimental data. But the most significant feature in Table~\ref{tab:outputs_fits} is the similar size achieved by the NLO coefficients, with values within the range $0.12-0.30$.

\begin{table}[h]
\centering
\begin{tabular}{lcc}
\hline \\[-2.5mm]
                                                     & WT+Born+NLO                                                      & WT+NLO+Born+RES   \\
\hline \\[-2.5mm]
$a_{\bar{K}N} \ (10^{-3})$           & $ 1.268^{+0.096}_{-0.096}$      & $ 1.517\pm 0.208$  \\
$a_{\pi\Lambda}\ (10^{-3})$        & $ -6.114^{+0.045}_{-0.055}$      & $ -2.624\pm 13.926$ \\
$a_{\pi\Sigma}\ (10^{-3})$           & $ 0.684^{+0.429}_{-0.572}$       &  $ 2.146 \pm 1.174$  \\
$a_{\eta\Lambda}\ (10^{-3})$      & $ -0.666^{+0.080}_{-0.140}$      &  $ 0.756\pm 1.215$ \\
$a_{\eta\Sigma}\ (10^{-3})$         & $ 8.004^{+2.282}_{-0.978}$      &  $ 10.105\pm 3.660$  \\
$a_{K\Xi}\ (10^{-3})$                   & $ -2.508^{+0.396}_{-0.297}$       &  $ -2.013\pm 0.743$ \\
$f/f_{\pi}$                                    & $ 1.196^{+0.013}_{-0.007}$        &  $ 1.180\pm 0.028$  \\
$b_0 \ (GeV^{-1}) $                      & $ 0.129^{+0.032}_{-0.032}$        &  $ -0.071\pm 0.016$  \\
$b_D \ (GeV^{-1}) $                      & $ 0.120^{+0.010}_{-0.009}$        &  $ 0.128\pm 0.015$   \\
$b_F \ (GeV^{-1}) $                       & $ 0.209^{+0.022}_{-0.026}$        &   $ 0.271\pm 0.022$  \\
$d_1 \ (GeV^{-1}) $                      & $ 0.151^{+0.021}_{-0.027}$        &   $ 0.144\pm 0.034$  \\
$d_2 \ (GeV^{-1}) $                      & $ 0.126^{+0.012}_{-0.009}$        &   $ 0.133\pm 0.011$  \\
$d_3 \ (GeV^{-1}) $                      & $ 0.299^{+0.020}_{-0.024}$        &   $ 0.405\pm 0.022$  \\
$d_4 \ (GeV^{-1}) $                     & $ 0.249^{+0.027}_{-0.033}$        &    $ 0.022\pm 0.020$  \\
$D$                                              & $ 0.700^{+0.064}_{-0.144}$         &    $ 0.700\pm 0.148$  \\
$F$                                               & $ 0.510^{+0.060}_{-0.050}$        &    $ 0.400\pm 0.110$  \\
$g_{\Lambda Y_{3/2} \eta} \cdot  g_{NY_{3/2}\bar{K}}$ &  \multicolumn{1}{c}{-}       & $ 8.924 \pm 11.790 $  \\
$g_{\Xi Y_{3/2}K} \cdot  g_{NY_{3/2}\bar{K}}$               &  \multicolumn{1}{c}{-}       & $ 6.200 \pm 8.214$  \\
$g_{\Xi Y_{5/2}K} \cdot  g_{NY_{5/2}\bar{K}}$               &  \multicolumn{1}{c}{-}       & $ -3.881 \pm 9.585 $  \\
$g_{\Xi Y_{7/2}K} \cdot g_{NY_{7/2}\bar{K}}$                &  \multicolumn{1}{c}{-}       & $ -14.306 \pm 14.427 $  \\
$\Lambda_{3/2}$ (MeV)                                                  &   \multicolumn{1}{c}{-}       & $ 839.66 \pm 406.68 $ \\                                                       
$\Lambda_{5/2}$ (MeV)                                                  &   \multicolumn{1}{c}{-}       & $ 541.31 \pm 290.01 $ \\
$\Lambda_{7/2}$ (MeV)                                                  &   \multicolumn{1}{c}{-}       & $ 500.00 \pm 426.82 $ \\
$M_{Y_{3/2}}$ (MeV)                                                      &   \multicolumn{1}{c}{-}       & $ 1910.00 \pm 44.70 $   \\
$M_{Y_{5/2}}$ (MeV)                                                       &   \multicolumn{1}{c}{-}      & $ 2210.00 \pm 39.07$   \\
$M_{Y_{7/2}}$ (MeV)                                                      &  \multicolumn{1}{c}{-}        & $ 2040.00 \pm 14.88 $    \\
$\Gamma_{3/2}$ (MeV)                                                   & \multicolumn{1}{c}{-}         & $ 200.00 \pm 120.31 $     \\
$\Gamma_{5/2}$ (MeV)                                                   & \multicolumn{1}{c}{-}         & $ 150.00 \pm 52.42 $     \\
$\Gamma_{7/2}$ (MeV)                                                   &   \multicolumn{1}{c}{-}       & $ 150.00 \pm 43.12 $     \\
\hline \\[-2.5mm]
$\chi^2_{\rm d.o.f.}$                                                       &  $1.14$                               &  $0.96$   \\
\hline
\end{tabular}
\caption{
Values of the parameters and the corresponding  $\chi^2_{\rm d.o.f.}$, defined in Eq.~(\ref{Chi^2_dof}), for the different fits described in the text. The subtraction constants are taken at a regularization scale $\mu=1$ GeV. The error bars in the parameters of the  WT+Born+NLO fit are determined as explained in the text, while those of the  WT+Born+NLO+RES fit are directly those provided by the MINUIT minimization procedure.}
\label{tab:outputs_fits}
\end{table}

\begin{figure*}[!ht]
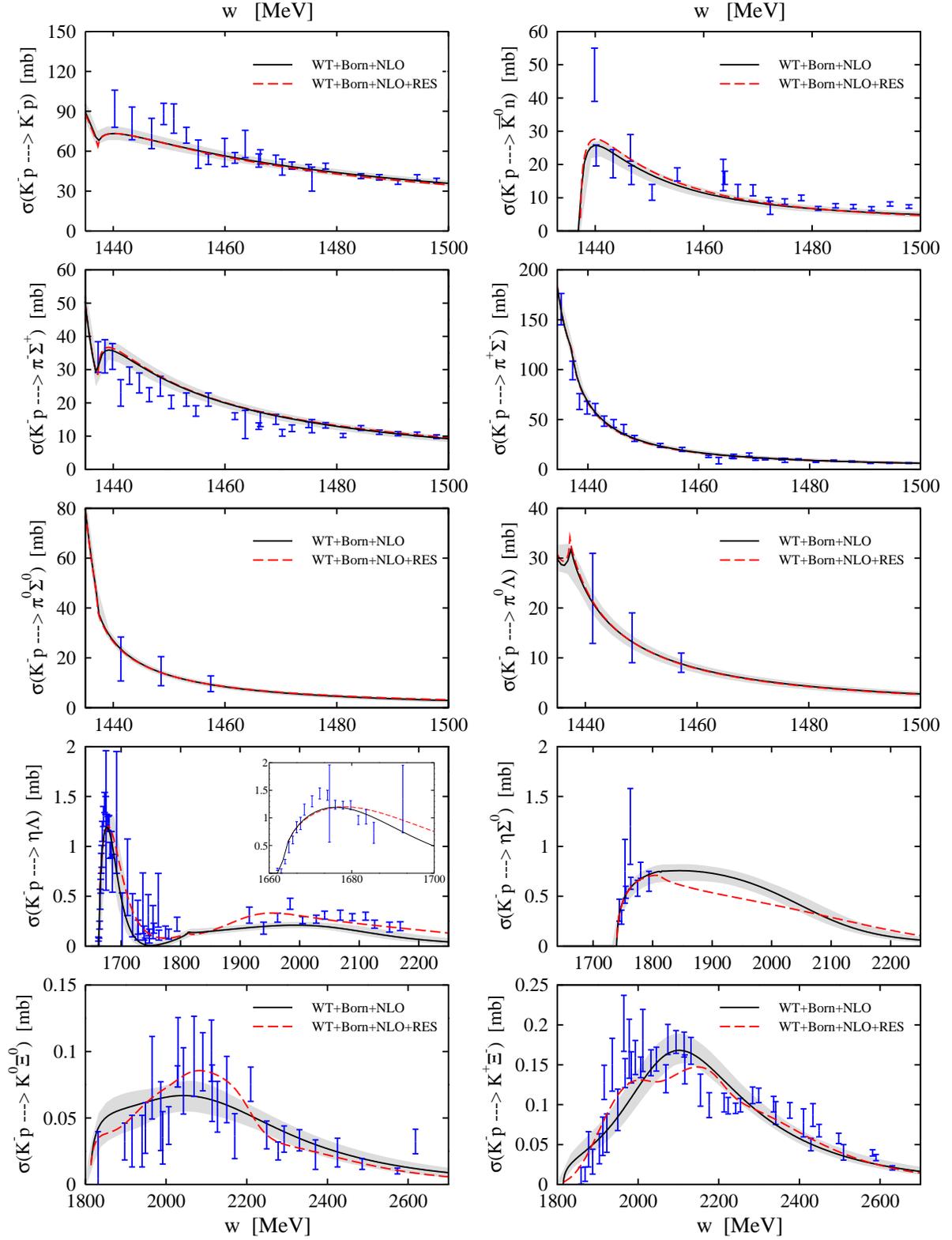

\centering
 \includegraphics[width=6.2 in]{All_chan.eps}
     \setlength{\unitlength}{0.01\linewidth}
    \begin{picture}(100,0)    
       \put(29.3,35.5){\includegraphics[width=18\unitlength]{etaL_zoom_thresh.eps}}
    \end{picture}
\caption{Total cross sections of the $K^- p\to K^- p, \bar{K}^0n, \pi^- \Sigma^+, \pi^+\Sigma^-, \pi^0 \Sigma^0, \pi^0\Lambda, \eta\Lambda, \eta\Sigma^0, K^+\Xi^-, K^0\Xi^0$ reactions obtained for the WT+Born+NLO fit (solid line), with the corresponding estimation of the error bands (grey area), and for the WT+Born+NLO+RES fit (dashed line). Experimental data have been taken from \cite{Starostin,Baxter,Jones,Berthon,exp_1,exp_2, exp_3, exp_4, exp_5, exp_6, exp_7, exp_8, exp_9, exp_10, exp_11}. See text for a detailed description of the models. The inset in the fourth panel of the left column shows the $K^-p \to \eta \Lambda$ cross-section in a reduced energy range close to threshold.} 
  \label{fig:xsect_all_chan}
\end{figure*}
\begin{table*}[!ht]
\centering
\begin{tabular}{lcccccc}
\hline \\[-2.5mm]
                             & {$\gamma$} & {$R_n$} & {$R_c$}&  {$a_p(K^-p \rightarrow K^- p)$}  & {$\Delta E_{1s}$}     & {$\Gamma_{1s}$} \\
\hline \\[-2.5mm]
Ikeda-Hyodo-Weise (NLO) \cite{IHW}   & 2.37   &  0.19    & 0.66    & $-0.70+{\rm i\,}0.89$                   & 306                          & 591  \\%
Guo-Oller (fit I + II) \cite{GO}  & $2.36^{+0.24}_{-0.23}$  & $0.188^{+0.028}_{-0.029}$  & $0.661^{+0.012}_{-0.011}$  & $(-0.69 \pm 0.16 ) + {\rm i\,}(0.94 \pm 0.11)$                   & $308 \pm 56$  & $619 \pm 73 $ \\%
Mizutani et al  (Model s) \cite{Mizutani:2012gy}   &  2.40  & 0.189    & 0.645    & $-0.69+{\rm i\,}0.89$  & 304  & 591      \\%
Mai-Meissner (fit 4) \cite{Mai:2014xna}  & $2.38^{+0.09}_{-0.10}$  &  $0.191^{+0.013}_{-0.017}$  & $0.667^{+0.006}_{-0.005}$  &    &  $288^{+34}_{-32}$   & $572^{+39}_{-38}$ \\%
Cieply-Smejkal (NLO) \cite{Cieply:2011nq}   &  2.37  & 0.191  & 0.660  & $-0.73+{\rm i\,}0.85$    & 310 & 607  \\%
Shevchenko (two-pole Model)  \cite{Shevchenko:2012np}  &     2.36     &    &      & $-0.74+{\rm i\,}0.90$  & 308   & 602 \\%
WT+Born+NLO             & $2.36^{+0.03}_{-0.03}$ & $0.188^{+0.010}_{-0.011}$ & $0.659^{+0.005}_{-0.002}$   & $-0.65^{+0.02}_{-0.08}+{\rm i\,}0.88^{+0.02}_{-0.05}$                   & $288^{+23}_{-8}$                         & $588^{+9}_{-40}$\\%
WT+NLO+Born+RES  & 2.36 & 0.189 & 0.661   & $-0.64+{\rm i\,}0.87$                   & 283                         & 587\\%
\hline \\[-2.5mm]
Exp.    & $2.36\pm 0.04$ &	$0.189\pm 0.015$ & $0.664\pm 0.011$        & $ (-0.66\pm0.07) + {\rm i\,}(0.81\pm0.15)$ &	$283\pm36$ & $541\pm92$ \\
\hline
\end{tabular}
\caption{Threshold observables obtained from our fits and from other recent studies \cite{IHW,GO,Mizutani:2012gy,Mai:2014xna,Cieply:2011nq,Shevchenko:2012np} which incorporated the SIDDHARTA measurements in the fitting procedure. Experimental data are taken from \cite{br_1,br_2,SIDD}. }
\label{tab:thresh}
\end{table*}

Even though the $\chi^2_{\rm d.o.f.}$ of the WT+Born+NLO fit (Table~\ref{tab:outputs_fits}) cannot be compared directly to that of the models in \cite{Feijoo:2015yja,Ramos:2016odk},  because we have included $58$ additional experimental points, one can check its goodness by looking at the agreement between experimental scattering data and the theoretical results present in Fig.~\ref{fig:xsect_all_chan}.

The total cross sections for $K^-p$ scattering to all channels of the $S=-1$ sector in the case of the WT+Born+NLO fit are represented by solid lines in Fig.~\ref{fig:xsect_all_chan},  while the grey area depicts the corresponding estimation of the error bands. Because of the novelty, we first inspect the total cross sections of the $\eta$ channels. One can clearly see that these cross sections are properly reproduced with this fit, excluding the small bump in the $\eta \Lambda$ cross section around $2000$~MeV where this theoretical model slightly underestimates its strength. The agreement with the experimental data just above the $\eta \Lambda$ threshold describing the $\Lambda(1670)$ resonant structure implies that this fit is able to dynamically generate such a resonance. 
It should be noted that the dynamical generation of this resonance was confirmed for the first time in \cite{Oset:2001cn} by means of a unitarized coupled-channels method using the lowest order (WT) chiral Lagrangian. 
The authors examined the contribution of the $\Lambda(1670)$ tail, and hence the role of the rescattering terms, on the $K^- p \to K \Xi$ reactions, because they found this resonance to couple strongly to $K \Xi$, being the squared value of the corresponding coupling one or two orders of magnitude larger than the ones to other isospin $0$ states in the $S=-1$ sector. Relatedly, the position of the pole associated to the $\Lambda(1670)$ was quite sensitive to the $a_{K \Xi}$ subtraction constant. The results obtained there clearly suggest that there is a correlation between the ability of a model in reproducing the $\Lambda(1670)$ resonance and the simultaneous accommodation of the $K \Xi$ production cross sections. This is a very valuable argument to discriminate among all possible parametrizations which describe in an acceptable way the $K \Xi$ cross sections. In this sense, the set of parameters of the present WT+Born+NLO fit is in line with the findings of \cite{Oset:2001cn}.

\begin{figure*}[ht]
\begin{center}
\centering
\includegraphics[width=5 in]{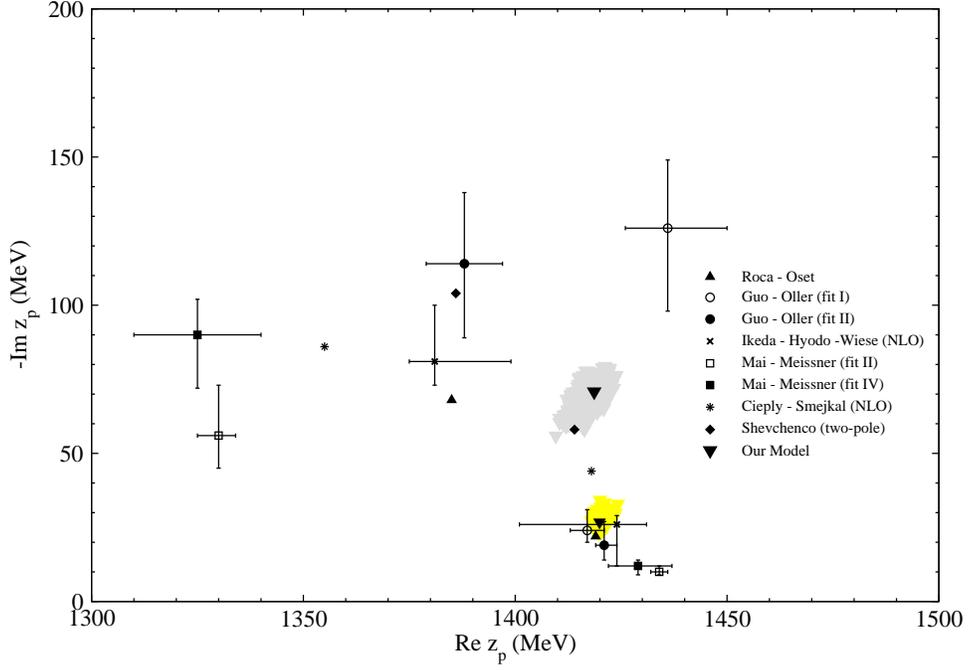}
\caption{\label{fig:pole_comparison} Pole positions of the $\Lambda(1405)$ for the approaches included in Table.~\ref{tab:thresh}, where we have also included the results from \cite{Roca:2013av}. The colored areas indicate the possible pole positions by varying the fitting parameters within their corresponding error bars in our WT+NLO+Born model. }
\end{center}
\end{figure*}

Concerning the $K^- p \to K^+ \Xi^-$ cross section (bottom panels in Fig.~\ref{fig:xsect_all_chan}), the WT+Born+NLO model gives a reasonable reproduction of data, although slightly worse than those obtained by our previous best pure chiral models \cite{Feijoo:2015yja,Ramos:2016odk}, since the maximum of the present distribution is shifted $50$~MeV towards higher energy. From \cite{Feijoo:2015yja,Ramos:2016odk}, one can also appreciate that the older models (WT+NLO and WT+NLO+Born respectively) clearly offer a better agreement with the experiment than the new one for the $K^- p \to K^0 \Xi^0$ cross section. 
We note that this is the price one has to pay in order to include and correctly describe the new channels $K^-p\to \eta\Lambda, \eta\Sigma$. Actually both older models, WT+NLO and WT+NLO+Born, miss the experimental data in these channels by up to an order of magnitude, as can be seen in \cite{Feijoo_thesis}. The difference in the behavior of the $K^- p \to K \Xi$ cross sections in the WT+Born+NLO model with respect to the older models, in particular the rather sharp rise of the $K^- p \to K^0 \Xi^0$ cross section just above the threshold, is related to the changed role of the isospin $I=0$ component, which becomes dominant at threshold energies as it picks up the tail of the dynamically generated $\Lambda(1670)$ resonance that describes the $K^- p \to \eta \Lambda$ reaction. Below we will discuss in detail the interplay between $I=0$ and $I=1$ components at different energies in the present and older models, showing in particular that our new model gives a much better prediction for the pure $I=1$ $K^0_L p\to  K^+ \Xi^0$ reaction. 

Coming back to the $K^- p \to K^0 \Xi^0$  and $K^- p \to K^+ \Xi^-$ cross sections, the model can be improved by the inclusion of resonant terms, similarly to what is done in \cite{Feijoo:2015yja}, where these proved to be very helpful to accommodate the theoretical cross section to the experimental data. We will discuss such a development in Sec.~\ref{subsec:res_inclusion}. However, looking at the bottom panels in Fig.~\ref{fig:xsect_all_chan}, we can already anticipate that the explicit inclusion of the $\Lambda(1890)$ is a good strategy, since it is located in the energy region of interest and it is an isospin $0$ resonance, the relevance of which is clarified at the end of this section.

Finally, the total cross sections of the classical processes obtained by the WT+Born+NLO fit, represented in the three top rows of Fig.~\ref{fig:xsect_all_chan}, reproduce the experimental data very well. This agreement is consequently reflected on the threshold observables, whose values are collected in Table~\ref{tab:thresh}, together with the results obtained by other works which include in their fits the recent experimental data from \cite{SIDD}. A similar degree of accuracy is reached by all the fits in reproducing the branching ratios, while, for the $K^-p$ scattering length and the related energy shift and width of the $1s$ state of kaonic hydrogen, the various models show slightly larger differences, yet all of them within the error range. These similarities can be attributed to their similar values of the $f$ parameter, given the dominance of the WT term at threshold, with the exception of the study \cite{Shevchenko:2012np}, which is based on a phenomenological strong isospin-dependent $K^-N - \pi\Sigma$ potential. 
 \begin{table}[h!]
\centering
\begin{tabular}{cccccc}

\hline \\ [-2.5mm]
\hline \\ [-2.5mm]
\multicolumn{6}{c}{ {\bf $0^- \oplus \frac{1}{2}^+$} interaction in {\bf$(I,S)=(0,-1)$} sector } \\ 
\hline \\[-2.5mm]
\multicolumn{2}{c}{Pole}  &$| g_{\pi \Sigma}|$  & $| g_{\bar{K}N}|$ &$| g_{\eta \Lambda}|$&$| g_{K \Xi}|$\\   
\hline \\[-2.5mm]
\multicolumn{2}{c}{$\Lambda(1405)$} & \multicolumn{4}{c}{} \\
\multicolumn{2}{c}{$1419^{+16}_{-22}-i~71^{+24}_{-31}$} & $3.40$ & $2.98$ & $1.10$ & $0.65$    \\
\multicolumn{2}{c}{$1420^{+15}_{-21}-i~27^{+18}_{-11}$} & $2.31$ & $3.51$ & $1.26$ & $0.36$    \\
\multicolumn{2}{c}{$\Lambda(1670)$} & \multicolumn{4}{c}{} \\
\multicolumn{2}{c}{$1675^{+10}_{-11}-i~31^{+4}_{-7}$} & $0.47$ & $0.59$ & $1.74$ & $3.71$    \\
\hline \\[-2.5mm]
 \multicolumn{6}{c}{ {\bf $0^- \oplus \frac{1}{2}^+$} interaction in {\bf$(I,S)=(1,-1)$} sector } \\ 
\hline \\[-2.5mm]
Pole               & $| g_{\pi \Lambda}|$ & $| g_{\pi \Sigma}|$  & $| g_{\bar{K}N}|$ & $| g_{\eta \Sigma}|$ & $| g_{K \Xi}|$ \\
\hline \\[-2.5mm]
$\Sigma^*$ & \multicolumn{5}{c}{} \\
$1701^{+16}_{-1}-i~170^{+2}_{-7}$ & $1.96$ & $0.47$ & $1.21$ & $0.36$ & $0.98$    \\
\hline \\ [-2.5mm]
\hline \\
\end{tabular}
\caption{
{\small  Pole positions of the pure chiral model WT+Born+NLO, expressed in MeV, and the corresponding modulus of the couplings $| g_i|$ found in isospin $0$ and $1$ basis.}}
\label{tab:spectroscopy}
\end{table}

We next analyze the pole content of the scattering amplitude derived from the WT+Born+NLO fit. 
The resonances found and their couplings to the different channels in the various sector are summarized in Table~\ref{tab:spectroscopy}.
In $I=0$, our model generates the double-pole structure of the $\Lambda(1405)$ and a pole representing the $\Lambda(1670)$. As can be appreciated in Table~\ref{tab:spectroscopy}, the broader pole of $\Lambda(1405)$, which couples mostly to $\pi\Sigma$, is located at $1419$~MeV. The other pole with smaller imaginary part is located very close to the former one, $1420$~MeV, but more strongly coupled to $\bar{K}N$. The characteristics of this narrower pole coincide with the results of other recent studies \cite{Roca:2013av,IHW,GO,Mizutani:2012gy,Mai:2014xna,Cieply:2011nq,Shevchenko:2012np}, which mostly fall within the range $(1415-1435, 10-25)$ MeV, as is reflected in Fig.~\ref{fig:pole_comparison} which contains the pole positions obtained by the previous authors in the complex plane. One also can see the huge differences in the position of the wider pole among the different groups. For a more detailed comparison between the pole content in some of the cited studies based on UChPT, we recommend the exhaustive analysis of \cite{Cieply:2016jby}. 
In Fig.~\ref{fig:pole_comparison}, we also include the dispersion of the pole positions for the configurations selected by the condition of Eq.~(\ref{chi_cond}) when we let the parameters vary within their error bands, as explained in Sec.~\ref{sec:Fit_Data_treat}. 

\begin{figure}[ht]
\begin{center}
\centering
\includegraphics[width=3.0 in]{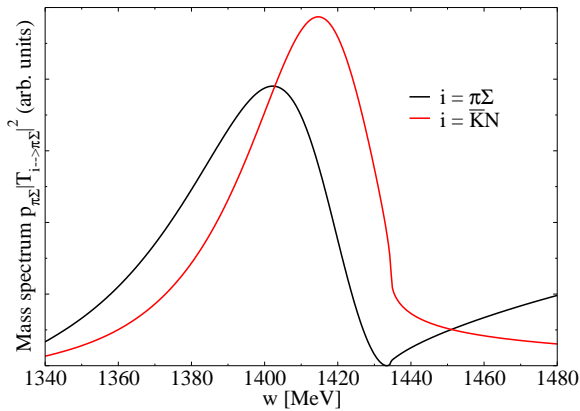}
\caption{\label{fig:SumT2_Lambda(1405)} Squared $I=0$ amplitudes for transitions from initial ${\bar K} N$ and $\pi \Sigma$ channels to final $\pi \Sigma$, weighted by the three-momentum of the final particles in the c.m.}
\end{center}
\end{figure}

Although we find the two poles to lie above the nominal value of the $\Lambda(1405)$ resonance, we note that the signal would be peaked between 1405 and 1415~MeV. This is clearly seen in Fig.~\ref{fig:SumT2_Lambda(1405)}, where we represent the quantities $q_{\pi\Sigma} \mid T_{\pi\Sigma \to \pi\Sigma} \mid^2$ and  $q_{\pi\Sigma} \mid T_{{\bar K} N \to \pi\Sigma} \mid^2$ as functions of the center-of-mass energy, with $q_{\pi\Sigma}$ being the momentum in the $\pi\Sigma$ center-of-mass frame. These quantities determine the cross section in experiments where the $\Lambda(1405)$ production is mainly driven by $\pi\Sigma$ or  ${\bar K}N$ intermediate states, respectively.

The $\Lambda(1670)$ resonance naturally shows up and helps describing the experimental $K^- p\to \eta \Lambda$ total cross sections. The corresponding properties are collected in Table~\ref{tab:spectroscopy}, from which we can appreciate that our results are in good agreement with \cite{PDG}, quoting a mass (1670-1680)~MeV and a width (25-50)~MeV. We note that, although within errors, our resonance is slightly wider than the average while its mass is completely coincident. The $\Lambda(1670)$ resonance found in the models of  \cite{GO} lies, on the contrary, on the narrow side of the error band, while the width of that found in \cite{Oset:2001cn} reproduces very satisfactorily the experimental one.

With respect to the $I=1$ resonances, we found a pole at $1701-i170$~MeV that can be related to some of the signals or bumps quoted by the PDG \cite{PDG} in the range 1600 to 1800~MeV.  An identification with the $\Sigma(1750)$ resonance is discarded, as this resonance is hardly seen to decay in $\pi\Lambda$ and has similar branching ratios to decay into ${\bar K}N$ and $\eta\Sigma$ states, a phenomenology that could not be reproduced with the coupling constants listed in Table~\ref{tab:spectroscopy}. However, our pole, which couples strongly to $\pi\Lambda$ states, could be associated to the two-star $\Sigma(1690)$ wide structure, which has been seen in production experiments only, and mainly in $\pi\Lambda$.

\subsection{Predictions for isospin 1 processes}
\label{subsec:JLab_prediction}

\begin{figure}[h!]
 \centering
  \includegraphics[width=3.0 in]{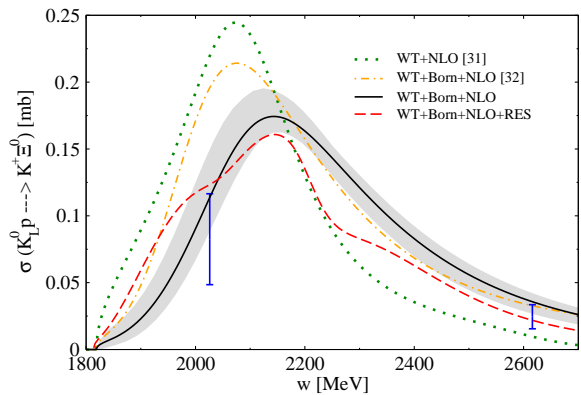}
 \caption{Total cross sections of the $K^0_L p\to  K^+ \Xi^0$ reactions for the for the models described in the present work (WT+Born+NLO and WT+Born+NLO+RES) and for the models of our previous studies \cite{Feijoo:2015yja,Ramos:2016odk}. A more detailed explanation can be found in the text. The grey area corresponds to the error band related to the WT+Born+NLO fit and the experimental points of the $I=1$ $K^- n \to  K^0 \Xi^-$ reaction, taken from  \cite{iso1exp1,iso1exp2} and divided by two, see Sect.~\ref{Isos_filters2} for more details.}
 \label{Jlab_prediction_for_all_models}
\end{figure}

In the Introduction and also in Sec.~\ref{Isos_filters2} we discussed the constraining effects of the inclusion of the future $K^0_L p\to  K^+ \Xi^0$ experimental data, because of its $I=1$ nature, on the NLO coefficients. To check the predictive power of the present models (WT+Born+NLO and WT+Born+NLO+RES), we present their corresponding $K^0_L p\to  K^+ \Xi^0$ cross section in Fig.~\ref{Jlab_prediction_for_all_models}. For completeness, we have also included the predictions of other previous models to see how the description of this observable evolves when more ingredients are taken into account. These two additional models \cite{Feijoo:2015yja,Ramos:2016odk} were fitted to the same experimental data collected in Table~\ref{tab_exp_points}, except for the total cross sections involving the $\eta$ channels, and their names inform on which terms are taken into account in the interaction kernel. Moreover, the experimental points of the pure $I=1$ $K^- n \to  K^0 \Xi^-$ reaction, which have been divided by $2$ to properly account for the size of the strangeness $S=-1$ component of the $K^0_L$ (see Sec.~\ref{Isos_filters2}), are also included in this figure. These two data points have not been used in any of the performed fitting procedures. 

As one can see from Fig.~\ref{Jlab_prediction_for_all_models}, the WT+Born+NLO (solid line), the WT+Born+NLO+RES (dashed line) and the WT+NLO+Born \cite{Ramos:2016odk} (dot-dashed line) fits do a good job at higher energy, while the  WT+NLO fit \cite{Feijoo:2015yja} (dotted line) provides a too low prediction. More significant is the behavior around $2$~GeV, where the old models overshoot the experimental cross section by about a factor of $2$, while the new models,  having their maximum strength shifted 50~MeV towards higher energies, provide a much better prediction.
We therefore see that, as more contributions are implemented in the interaction kernel and more data are included in the fits, especially from isospin filtering processes, the results from the theoretical models get closer to the available experimental points. Given these results, everything seems to indicate that new data from scattering induced by the secondary $K^0_L$ beam proposed at Jlab would be very helpful to further constrain the theoretical models.
 
Another opportunity to check the validity of our models has been recently provided by the AMADEUS Collaboration, which has determined
the $I=1$ $K^- p \to \pi^- \Lambda$ amplitude around threshold \cite{Piscicchia:2018rez} from $K^-$ absorption processes on $^4$He. The chiral WT+Born+NLO model predicts a value of $0.38^{+0.02}_{-0.06}$~fm, while the model with resonances WT+Born+NLO+RES gives  0.36~fm. These values are
in complete agreement with the experimental determination of  $\mid A_{K^- n \to \pi^- \Lambda}\mid = (0.334\pm 0.018~{\rm stat}^{+0.034}_{-0.058}~{\rm syst})$~fm.

\subsection{$\Lambda_b \to J/\Psi ~ K\Xi, ~ J/\Psi ~ \eta \Lambda$ prediction}
\label{subsec:Lambda_b_prediction}
\begin{figure*}[!htb]
\centering
  \includegraphics[width=4.5in]{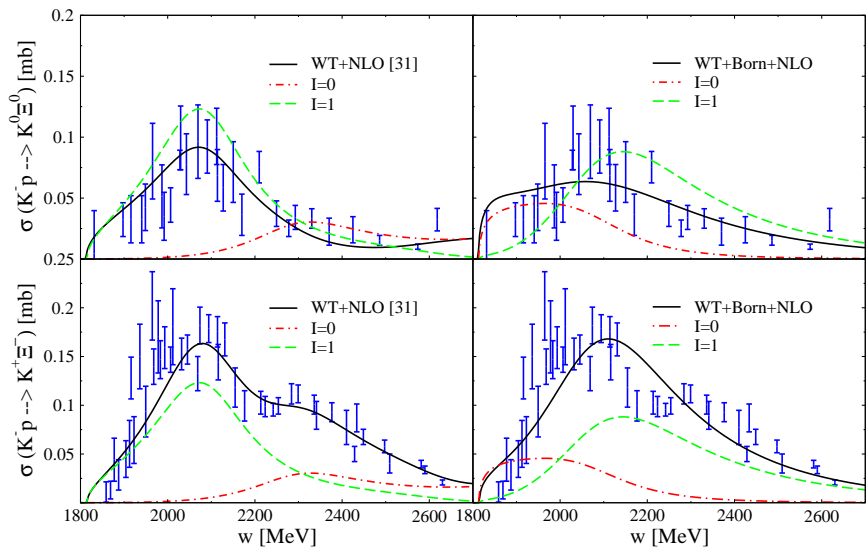}
\caption{The total cross section data of the  $K^- p \to K^0 \Xi^0$ reaction is represented in the top panels, where the left figure corresponds to the WT+NLO model \cite{Feijoo:2015yja} and the right one corresponds to the WT+Born+NLO model. The same distribution for the bottom panels where the $K^- p \to K^+ \Xi^-$ cross section data are represented. The figure shows the complete results by means of solid lines, the results where only isospin $I=1$ component (dashed lines) or the $I=0$ one (dot-dashed line) have been retained.}
\label{isospin_distr}
\end{figure*}

We start this subsection by presenting in Fig.~\ref{isospin_distr} the cross section of the $K^- p\to K^0 \Xi^0$ reaction (top panels) and the $K^- p\to K^- \Xi^+$ reaction (bottom panels), obtained from the old WT+NLO \cite{Feijoo:2015yja} (left panels) and the new WT+Born+NLO (right panels) models. Let us remind the reader that the WT+NLO model employs the dynamics of the chiral Lagrangian up to NLO, specifically the contributions of the WT term and the NLO ones. The WT+Born+NLO model also considers the Born terms and includes additional data in the fits, namely the cross sections of the $ K^- p \to \eta \Lambda, \eta \Sigma $ processes. The WT+NLO model was employed in our first study \cite{Feijoo:2015cca} of the $\Lambda_b$ decays into $J/\Psi K\Xi$ and $ J/\Psi  \eta \Lambda$ states and we want to compare with the predictions of our new fit in the present work. The figure shows the complete cross sections (solid lines), as well as the results obtained when only the isospin $I=1$ component (dashed lines) or the $I=0$ one (dash-dotted lines) are retained. It is interesting to see that, in both models, the $I=1$ component is dominant. The $I=1$ distribution for the WT+NLO model is a little bit more enhanced at low energies than that corresponding to the WT+Born+NLO model that peaks about 50~MeV higher away in energy. The contribution of $I=0$ to the $K^-p \to K\Xi$ cross section for the WT+NLO model is mainly significant around  and beyond $2300$~MeV, which is in contrast to what is found for the WT+Born+NLO model: a $ I = 0 $ distribution that grows rapidly near threshold and, after reaching a plateau, experiences a smooth fall, being practically negligible around 2300~MeV.

\begin{figure}[!htb]
\centering
  \includegraphics[width=2.8in]{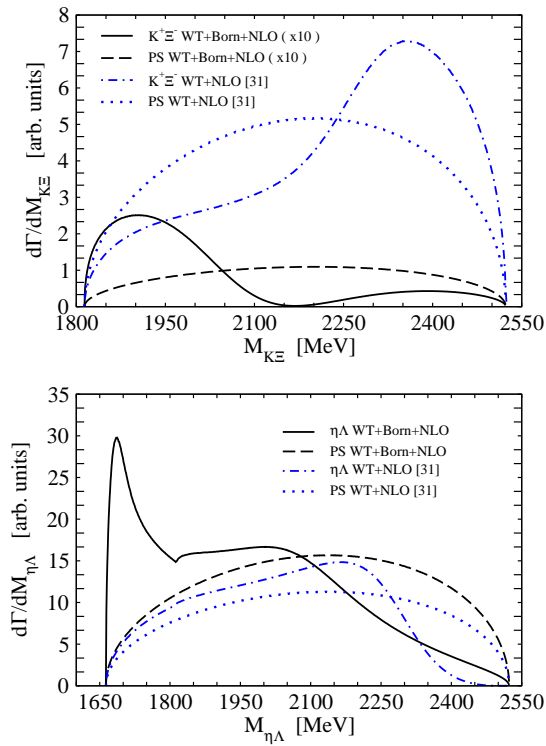}
\caption{ Invariant mass distributions, in arbitrary units, of $K^+\Xi^-$ states (upper panel) and $\eta \Lambda$ states (lower panel) obtained for the two models discussed in this section: WT+NLO \cite{Feijoo:2015yja} (dash-dotted lines) and WT+Born+NLO (solid lines). The pure phase-space (PS) distributions (dotted and dashed lines) are normalized to the corresponding invariant mass distribution; see more details in the text. The invariant mass distribution of $K^+\Xi^-$ state and the corresponding phase-space distribution for WT+Born+NLO are multiplied by a factor 10 to aid its visualization.}
  \label{fig:PS}
\end{figure}

In the upper panel of Fig.~\ref{fig:PS} we present the invariant mass distributions of $K^+\Xi^-$ pairs from the decay process $\Lambda_b \to J/\psi ~ K^+ \Xi^-$. We note that the invariant mass distribution of $K^+\Xi^-$ states and the corresponding phase-space distribution for the WT+Born+NLO model are multiplied by a factor $10$ to aid its visualization. The fact that this decay filters the $I=0$ components makes the differences between the WT+NLO (dash-dotted line) and WT+Born+NLO (solid line) models more evident, not only in the strength but also in the shape of the invariant mass distribution, which is in accordance with the $I=0$ cross sections shown in Fig.~\ref{isospin_distr}. The strength from the WT+NLO model exceeds by more than a factor 20 that from the WT+Born+NLO model. The reason lies in the fact that the $\Lambda_b$ decay processes producing the $J/\psi$  do not allow the formation of primary $K\Xi$ pairs, which are only produced through rescattering of the $\bar{K}N$ and $\eta \Lambda$ primary components. Thus, the $\Lambda_b \to J/\psi~ K^+\Xi^-$ reaction is directly proportional to the meson-baryon scattering amplitude, specifically to the $\eta \Lambda \to K \Xi$ and $\bar{K}N \to K \Xi$ components in $I=0$, which can lead to a marked interference pattern. Indeed, the invariant mass distribution of the $K^+\Xi^-$ state for the WT+Born+NLO model is a clear example of a strong destructive interference among the  $\bar{K}N$ the $\eta \Lambda$ mediated processes.
By contrast, the $\Lambda_b \to J/\psi ~\eta \Lambda$ decay can proceed at tree level, making the possible interference effects of the loop diagrams acquire a secondary role, which explains the similar size of the distributions for this process obtained by the two models. Note that, in spite of the lack of a direct term in the $\Lambda_b \to J/\psi~ K^+\Xi^-$  decay, its strength in the case of the WT+NLO model is comparable to that of the $\Lambda_b \to J/\psi ~\eta \Lambda$ decay, indicating a strong constructive interference of the rescattering terms in the former reaction, just the opposite of what happens for the WT+Born+NLO model. 

The phase space (PS) distributions shown in Fig.~\ref{fig:PS} permit a comparison with the corresponding invariant mass distributions to point out the dynamical features in the meson-baryon amplitudes. The PS distributions are obtained by taking the amplitude ${\cal M}$ as a constant in Eq.~(\ref{eq:double_diff_cross}) and normalizing to the area of the invariant mass distribution of the corresponding model. In the case of the WT+NLO model, we observe a peaked structure around $2350$~MeV in the $ K^+\Xi^-$ distribution. This is not a resonance but merely a reflection of the phase space limitation at about $2500$~MeV which produces a narrower structure than that in the cross sections of the $K^-p \to K \Xi$ reactions, as we can see from the much broader $I=0$ contribution in Fig.~\ref{isospin_distr} (left panels). Actually, if it was a resonance, a peaked structure would also appear in the $\eta \Lambda$ mass distribution at the same energy, which is absent. An analogous explanation holds for the bump observed in the invariant $ K^+\Xi^-$ mass distribution in the case of the WT+Born+NLO  model at the lower end of the spectrum. On the other hand, the $ \eta \Lambda $ invariant mass distribution corresponding to the WT+Born+NLO model makes evident the presence of the dynamically generated $\Lambda(1670)$ resonance. What is clear from this study is that the measurement of these $\Lambda_b$ decay channels will provide valuable information concerning the meson-baryon interaction at higher energies, beyond what is offered to us by present scattering data.

\subsection{Inclusion of resonances: WT+Born+NLO+RES fit}
\label{subsec:res_inclusion} 

As already mentioned, the main motivation for carrying out a new fit including the resonances is to study the implications of these terms on the low-energy constants. As in the previous fit, the inclusion of data at higher energies has not affected the quality of the low-energy observables. A clear proof of it is the good description of the low-energy data compiled in Table~\ref{tab:thresh} and in the six upper panels in Fig.~\ref{fig:xsect_all_chan}. The inclusion of resonances in the new WT+Born+NLO+RES model improves slightly the overall agreement of the threshold observables, as we see in Table~\ref{tab:thresh}. The total $K^- p\to K^- p, \bar{K}^0n, \pi^- \Sigma^+, \pi^+\Sigma^-, \pi^0 \Sigma^0, \pi^0\Lambda $ cross sections obtained by the WT+Born+NLO+RES model (dashed line in Fig.~\ref{fig:xsect_all_chan}) offer a very similar description, almost indistinguishable to naked eye, to that of the  WT+Born+NLO one.

Obviously, much more pronounced structures in the cross sections are expected to be provided by the additional resonant contributions in the processes whose amplitudes contain explicitly such terms, namely the $K^- p \to \eta \Lambda, K^0 \Xi^0, K^+ \Xi^- $ ones. Partly, the inclusion of the $\Lambda(1890)$ resonance motivated by the lack of agreement between the WT+Born+NLO model (solid line in the four bottom panels of Fig.~\ref{fig:xsect_all_chan}) and the scattering data corresponding to the $K^- p \to K^0 \Xi^0$ reaction at low energy and to the $K^- p \to \eta \Lambda$ reaction around $1950$~MeV. From the dashed lines in the corresponding panels, one can appreciate a clear improvement in reproducing the experimental $K^- p \to \eta \Lambda$ cross section in the energies ranging from $1850$ to $2200$~MeV without affecting the resonant structure from the $\Lambda(1670)$. But, probably, the most notable effect is the one observed in the results for the $K^- p \to K^0 \Xi^0$ cross section, which now reproduces better the experimental data just above threshold. The reason becomes evident when looking at the distribution of the $I=0$ component of the $K\Xi$ production of WT+Born+NLO in Fig.~\ref{isospin_distr} (right panels), which shows that $\Lambda(1890)$ resonance has a prominent background to interfere with. The favorable contribution of the $\Sigma(2030)$ and $\Sigma(2250)$ resonances can also be appreciated in the $K^- p \to K^0 \Xi^0, K^+ \Xi^-$ cross sections shown in Fig.~\ref{fig:xsect_all_chan}. Again, this is due to the interference of these resonant terms with the $I=1$ distribution seen in Fig.~\ref{isospin_distr} (right panels), which introduces some structure and provides a better account of the experimental data in the $2000-2200$~MeV energy range.
With respect to the $K^- p \to \eta \Sigma^0$ cross section, we do not notice any difference in the reproduction of the experimental data, but for energies around $1800$~MeV the WT+Born+NLO+RES model presents a more pronounced slope. In summary, this better general description of the experimental data results in a 16\% of improvement in the goodness of the fit, as reflected by the  $\chi^2_{\rm d.o.f.}$ values in Table~\ref{tab:outputs_fits}.

Now we turn our attention to the analysis of the fitting-parameter stability, which was the main goal when performing the WT+Born+NLO+RES fit. From Table~\ref{tab:outputs_fits}, it can be appreciated that the fitting parameters are quite stable when going from the purely chiral-model fit to the one including resonances, the reason stemming in the fact that
we have employed additional observables that are sensitive to the NLO term, namely the scattering data from the $K^- p\to \eta \Lambda$  reaction (apart from $ K^0 \Xi^0, K^+ \Xi^- $ reactions). In other words, although the inclusion of resonances help in fine-tuning the agreement with experimental data, the so-called background contributions obtained by the WT+Born+NLO model already perform an excellent job.  Going more into the details, we would like to first stress the particular case of the $d_2$ and $d_3$ coefficients, which have kept especially stable even from our earlier model, WT+NLO+Born \cite{Ramos:2016odk}, which does not employ the $K^-p \to \eta\Lambda, \eta\Sigma$ data. This is tied to the prominent role that these NLO coefficients have in the description of the $ K^0 \Xi^0, K^+ \Xi^- $ reactions, as can be seen by exploring the $L_{ij}$ Clebsch-Gordan-type coefficients for these transitions in \cite{Feijoo:2015yja}. When including the resonances, the parameter $d_3$ varies more prominently, but still within a moderate 30\%. 
The stability of other parameters has also clearly increased by the inclusion of the $K^-p \to \eta\Lambda,\eta\Sigma$ reactions. The additional parameters relevant for the NLO  $K^- p \to \eta \Lambda,\eta\Sigma$ amplitudes, are $b_D$, $b_F$ and $d_1$, which appear in the corresponding $D_{ij}$ and $L_{ij}$ Clebsch-Gordan-type coefficients \cite{Feijoo:2015yja}. The stability of these parameters can be noticed when comparing the outputs of the WT+Born+NLO and WT+Born+NLO+RES models in Table~\ref{tab:outputs_fits}. This is in contrast with the large volatility of the $b_0$ and $d_4$ parameters, which only have a role in the elastic transitions of the sector as can be seen from Table~VIII in \cite{Feijoo:2015yja}, thereby not being directly affected by the experimental data employed here. Measuring processes that are affected by elastic $S=-1$ meson-baryon amplitudes that cannot proceed at lowest order, such as the decay $\Lambda_b \to J/\Psi ~ \eta \Lambda$ studied here involving the $t_{\eta\Lambda,\eta\Lambda}$ diagonal transition, would provide valuable information to better constrain the $b_0$ and $d_4$ parameters of the NLO Lagrangian.
As a consequence of the overall stability of the NLO parameters, the WT+Born+NLO+RES model also produces natural sized subtraction constants. 

Another interesting result shown in Table~\ref{tab:outputs_fits} is that the $f$ parameter has decreased after the inclusion of the resonances, which is in contrast to what happens when comparing the $f$ values for the different models in \cite{Feijoo:2015yja}, where this parameter remains almost invariable.

\section{CONCLUSIONS}
\label{sec:conclusions}
 
We have performed a new study of the meson-baryon interaction in the $S=-1$ sector in s-wave, aiming at an improved determination of the NLO terms of the chiral SU(3) Lagrangian that effectively permits the description of the data in a wide energy range.  To this end, we have taken into account in the fitting procedure experimental data of processes that are especially sensitive to higher orders of the Lagrangian, such as the $K^- p\to K^+\Xi^-, K^0\Xi^0$ reactions. We have also included explicit resonant contributions in order to test the stability of the NLO order parameters. 

The novelty with respect to our previous studies is the 
consideration in the present work of additional isospin filtering reactions, $K^- p \to \eta \Lambda$ in $I=0$ and $K^- p \to \eta \Sigma^0$ in $I=1$. The former reaction is interesting because it is also sensitive to the NLO terms.
Our purely chiral fit describes the $\Lambda(1670)$ resonant structure just above the $\eta\Lambda$ threshold, indicating the dynamical origin of such state, as found in other works in the literature. We also find this resonance to couple strongly to $K\Xi$ states. Therefore, even if located subthreshold, its tail influences, through rescattering, the $K^-p\to K\Xi$ cross sections studied here. 
The most remarkable feature of the incorporation of the $K^-p \to \eta\Lambda,\eta\Sigma^0$ data in the fits is the homogeneity of values found for the next-to-leading order low-energy constants, which, in turn, also show to be of rather natural size.

The inclusion of explicit resonances in the $K^-p\to K^+\Xi^-$, $K^0\Xi^0$ and  $\eta \Lambda$ amplitudes incorporates higher-angular momentum contributions and does improve the description of the corresponding cross sections in the 2~GeV region, as expected. We find that the fitted parameters remain quite stable when going from the purely chiral-model fit to the one including resonances, indicating that the former model already implements sufficient observables that 
are sensitive to the chiral NLO terms and the effect of higher angular momentum components is moderate.
In other words, although
 the inclusion of resonances help in fine-tuning the agreement with experimental data, the so-called background contributions obtained by the pure chiral model already perform satisfactorily. In particular, we find
the coefficients $d_2$ and $d_3$ to be very sensitive to the  $K^- p \to K^0 \Xi^0, K^+ \Xi^- $ cross sections, while $b_D$, $b_F$ and $d_1$ play a prominent role in the $K^-p \to \eta \Lambda,\eta\Sigma$ ones. Only the $b_0$ and $d_4$ parameters are not directly affected by the experimental data employed here, which explains their lower degree of stability.

We have also given predictions for new single isospin component processes, such as the $I=1$ $K^0_L p \to K^+ \Xi^0$ reaction that could be measured in the proposed secondary $K^0_L$ beam at Jlab. We have observed that, as more contributions are implemented in the interaction kernel and more data are included in the fits, especially from isospin filtering processes, the predictions of the model get much closer to the two available experimental points extracted from the $K^- n \to  K^0 \Xi^-$ reaction. 
Our result for the strength of the $I=1$ $K^- p \to \pi\Lambda$ amplitude at energies close to threshold is also in complete agreement with the recent determination provided by the AMADEUS Collaboration from $K^-$ absorption processes on $^4$He. The 
weak decay of the $\Lambda_b$  into a $J/\Psi$ and different meson-baryon pairs selects then the $I=0$ component of their interaction and is available at LHCb. We have presented results for the decay of the $\Lambda_b$  into $J/\Psi K^+ \Xi^-$  and $J/\Psi \eta \Lambda$ states and have found the invariant mass distributions to depend strongly on the meson-baryon interaction model employed. 

In summary, our study has made evident the success of isospin filtering processes in constraining the parameters of the chiral Lagrangian. The measurement of the above mentioned reactions, that select a pure isospin component,  would greatly contribute to having a better knowledge of the $S=-1$ meson-baryon interaction.

\section*{ACKNOWLEDGMENTS}

This work is partly supported by the Spanish Ministerio de Economia y Competitividad (MINECO) under the project MDM-2014-0369 of ICCUB (Unidad de Excelencia 'Mar\'\i a de Maeztu'), 
and, with additional European FEDER funds, under the contracts FIS2014-54762-P and
FIS2017-87534-P.


\end{document}